\documentclass[]{aa} 

\usepackage{graphicx}
\usepackage{txfonts}
\usepackage{xcolor}

\usepackage[breaklinks, colorlinks, citecolor=blue, linkcolor=blue]{hyperref}

\graphicspath{{/home/aasensio/Dropbox/GIT/DeepLearning/spatially\_variant/validate/figs}}

\newcommand{\vect}[1]{\boldsymbol{\mathbf{#1}}}

\begin{document} 

   \title{Solar multi-object multi-frame blind deconvolution\\ with a spatially variant convolution neural emulator}

   \author{A. Asensio Ramos}

   \institute{Instituto de Astrof\'isica de Canarias (IAC), Avda V\'ia L\'actea S/N,
		38200 La Laguna, Tenerife, Spain\\
              \email{andres.asensio@iac.es}
         \and
             Departamento de Astrof\'isica, Universidad de La Laguna, 38205 La Laguna, Tenerife, Spain\\
              }

   \date{Received ; accepted }

 
  \abstract
   {The study of astronomical phenomena through ground-based observations is always challenged 
   by the distorting effects of Earth's atmosphere. Traditional methods of post-facto image correction, 
   essential for correcting these distortions, often rely on simplifying assumptions that limit 
   their effectiveness, particularly in the presence of spatially variant atmospheric turbulence. Such
   cases are often solved by partitioning the field-of-view into small patches, deconvolving each patch
   independently, and merging all patches together. This approach is often inefficient and
   can produce artifacts.}
   {Recent advancements in computational techniques and the advent of deep learning offer new pathways 
   to address these limitations. This paper introduces a novel framework leveraging a deep neural 
   network to emulate spatially variant convolutions, offering a breakthrough in the efficiency and 
   accuracy of astronomical image deconvolution.}
   {By training on a dataset of images convolved with 
   spatially invariant point spread functions and validating its generalizability to spatially variant conditions, 
   this approach presents a significant advancement over traditional methods. 
   The convolution emulator is used as a forward model in a multi-object multi-frame blind deconvolution
   algorithm for solar images.}
   {The emulator enables the deconvolution of solar observations across large 
   fields of view without resorting to patch-wise mosaicking, thus avoiding artifacts associated 
   with such techniques. This method represents a significant computational advantage, reducing 
   processing times by orders of magnitude.}
   {}


   \maketitle
%
\section{Introduction}
\label{sec:introduction}
Ground based observations are always affected by the perturba-
tions produced by the atmosphere.
The turbulence induced by unavoidable air motions (seeing) produces changes in the
refraction index at many spatial and temporal scales. These changes affect the incoming light from astronomical
objects by producing local deviations. The phase
of the incoming wavefront, which can be safely assumed to be flat outside the
Earth atmosphere, is corrugated by the presence of atmospheric turbulence. This produces, in
general, a blurring effect that is absent in observations from observatories
located in space.

The temporal and spatial characteristics of the turbulence change with time and
they are difficult to predict. The temporal correlation time of the atmosphere
is typically very short, close to milliseconds. As a consequence, 
observations with long integration times (longer than the correlation time) average
the state of the atmosphere over many realizations. In such a case, due to the information
loss, the blurring effect of the atmosphere is unavoidable and cannot
be corrected for. On the other hand, observations with very short 
integration times (of the order of the correlation time) freeze the state of 
the atmosphere and can be used to reconstruct the original object from the information
present in the speckle distribution. Short integration 
times are, though, affected by noise and many such observations are required to reduce its
impact in the final result. 
This is specially relevant in low luminosity observations (e.g., observing
exoplanets, faint stars, etc.) or when doing high spectral resolution spectro-polarimetry.
Concerning the spatial correlation, strong turbulence produces eddies of very short
size (of the order of cm). As a consequence, the ensuing wavefront is very corrugated by the
presence of many short scale eddies covering the projection of the pupil of the telescope on
the sky. This is especially relevant for large aperture telescopes, where the size of the pupil is large.
To make things more challenging the turbulence existing
at different heights in the atmosphere is potentially
different. Turbulence close to the pupil of the telescope (low-height
turbulence) produces the same blurring in the whole field-of-view (FoV). 
Meanwhile, turbulence at higher altitudes produces differential blurring
in different parts of the FoV, known as anisoplanatism.

Several techniques have been developed for reducing the perturbing effect of the 
atmosphere. Adaptive optics (AO) is arguably one of the most succesful techniques. This
technique is based on the measurement of the instantaneous wavefront (ideally
at millisecond cadence) and compensate for the measured phase by using
active optics like deformable mirrors. Ground layer AO, which measure the 
wavefront in the pupil of the telescope and correct for it using deformable
mirrors produce an improvement of the image quality in the whole
FoV. However, the presence of turbulence at high atmospheric
layers that produce differential effects in different parts of the FoV requires
the application of multi-conjugate AO systems. These systems are in charge of measuring 
the turbulence at many atmospheric layers on real time and correcting for them using 
deformable mirrors conjugated at different heights in the atmosphere. These systems are 
still in their infancy and they are very complex to operate in real time.

Another very succesful technique for reducing the effect of the atmosphere is
post-facto image reconstruction. These methods are based on using a physical model
for the formation of an image under a turbulent atmosphere and inferring
the object and the instantaneous status of the atmosphere applying a maximum
likelihood (ML) approach. This requires the observation of several short
exposure time observations. Better results are obtained when using a relatively
large number of observations (of the order of tens for typical solar
observations). The problem to be solved is of the blind deconvolution kind,
since neither the object nor the instantaneous wavefronts are known and need to be inferred
simultaneously. Like any blind deconvolution problem, they are severly ill-defined and some
type of regularization is required. In the literature one can find methods 
based on maximum a-posteriori, which make use of the Bayes theorem and
explicit priors both for the object and the wavefronts \citep[see][and references therein]{molina01}. Other
methods use the simpler ML approach but introduce regularization by using a low complexity description of the
wavefront and making use of spatial filtering to avoid unphysical high frequencies to
appear in the deconvolved image \citep{lofdahl_scharmer94,vannoort05}.

\begin{figure*}
  \centering
  \sidecaption
  \includegraphics[width=12cm]{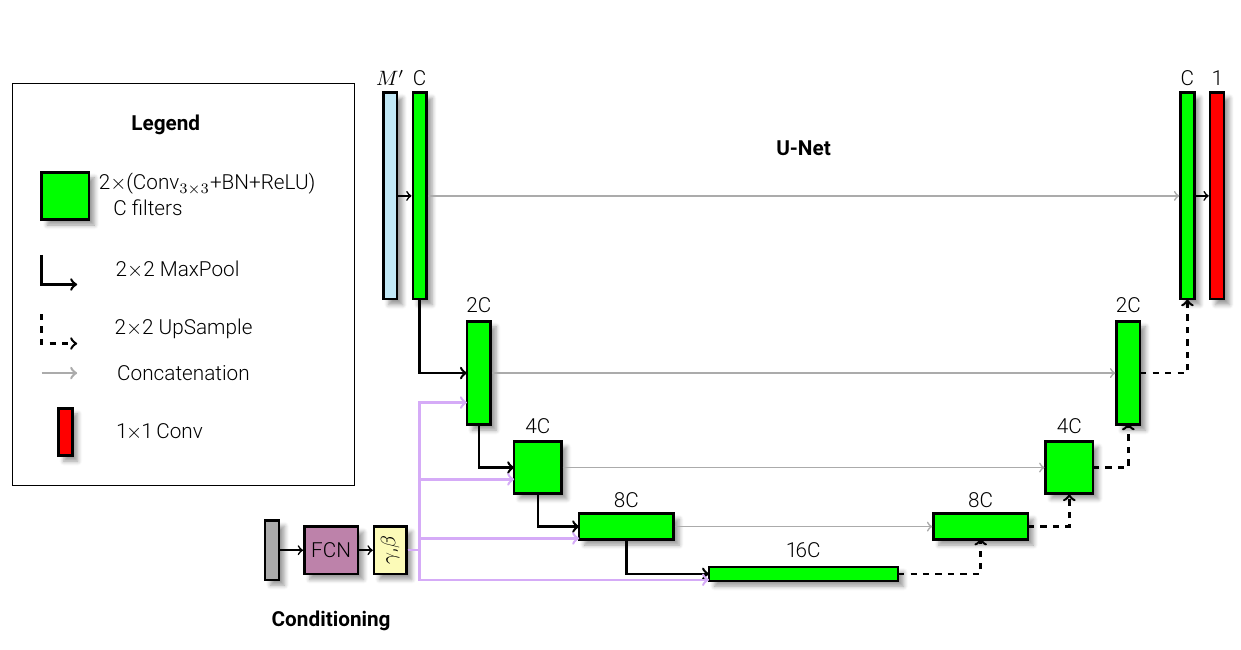}
  \caption{U-Net deep neural network used in this work as an emulator for
  the spatially variant convolution. $M'=M+1$ is the number of channels
  of the input tensor, that is built by concatening the image of 
  interest and the spatially variant images of the $M$ Karhunen-Lo\`eve.
  All green blocks of the encoder are conditioned to the specific 
  instrumental configuration as described in Sect. \ref{sec:encoder} via
  the encoding fully connected network shown in the figure. The grey 
  box refers to the one-hot encoding of the instrumental configuration.}
  \label{fig:unet}
\end{figure*}

The most widespread physical model used for blind deconvolution requires the assumption
of a spatially invariant point spread function (PSF). Its main advantage is that the
convolution of the object with the PSF can be computed very efficiently using the fast
Fourier transform (FFT). As a consequence, the
deconvolution can only be carried out inside one anisoplanatic patch,
where this assumption holds. This assumption is 
somewhat restrictive under the presence of strong turbulence or when
observing with a large aperture telescope. In those two cases, the anisoplanatic
patch becomes very small. For overcoming this limitation, arguably the most 
widespread solution is the application of the overlap-add (OLA) method. This method
requires mosaicking the image in
sufficiently small overlapping patches, deconvolving each patch 
independently under the assumption of a spatially invariant PSF, and finally
merging all patches together. This merging operation is commonly performed by
averaging each patch with a weighting scheme to smoothly connect
the overlapping borders. This approach has been very succesful in 
solar observations, as demonstrated by the success of the multiframe multiobject
blind deconvolution code \citep[MOMFBD;][]{vannoort05}. More elaborate
and precise methods do however exist to deal with spatially variant
PSFs, like the widespread method of \cite{nagy_oleary98} and
the recent space-variant OLA \citep{hirsch10}. A good
review of these techniques can be found in \cite{denis15}.

We propose in this paper a novel approach to deal with spatially variant PSFs. We propose to
use a deep neural network to emulate the spatially variant convolution with
a PSF that can be arbitrarily different in all pixels in the FoV. 
We check that this spatially-variant convolution emulator (SVCE), although trained 
with spatially invariant convolutions, 
correctly approximates a spatially variant convolution with good 
precision. The deep learning model is fully convolutional, so that it
generalizes to images of arbitrary size. We demonstrate the use of
this SVCE as a very efficient forward model to carry out multiobject multiframe blind 
deconvolutions in large FoV without using any mosaicking.

\section{Neural spatially variant convolution}
\label{sec:spatially_variant}
In the general case in which the PSF of an optical system (the atmosphere plus
the telescope and instruments in our case) is spatially variant, 
the intensity $I(\mathbf{r})$ received at a certain pixel location defined by the 
coordinates $\mathbf{r}=(x,y)$ is given by the following integral in the auxiliary
variable $\mathbf{r}'$:
\begin{equation}
  I(\mathbf{r}) = \int O(\mathbf{r}') S(\mathbf{r},\mathbf{r}') \mathrm{d}\mathbf{r}',
\end{equation}
where $O(\mathbf{r}')$ is the object intensity before the optical system 
and $S(\mathbf{r},\mathbf{r}')$ is the local PSF at point $\mathbf{r}$. Computing the measured intensity for a 
pixellized image of size $N \times N$ requires the computation of $N^2$ such
integrals. Therefore, the computation of the final image scales as 
$\mathcal{O}(N^4)$, which is prohibitive for large images. One way of 
reducing the computing time is by assuming a relatively compact PSF, so that it can be 
correctly defined in a pixellized image of size $M \times M$, with 
$M \ll N$. In such case, the computing work scales as $\mathcal{O}(N^2 M^2)$, which can
still be prohibitive for large images.

Contrarily, when the PSF is spatially invariant, $S(\mathbf{r},\mathbf{r}')=S(\mathbf{r}-\mathbf{r}')$ 
is fulfilled, and the integral can be computed by using the convolution theorem:
\begin{equation}
  I(\mathbf{r}) = O(\mathbf{r}) \otimes S(\mathbf{r}).
\end{equation}
From a computational point of view, the spatially invariant case can be
very efficiently computed using the FFT, since
the FFT of a 2D image can be computed in $\mathcal{O}(N^2 \log N)$ operations.
Three FFTs are required to compute the convolution, so that the total
computing time scales as $\mathcal{O}(N^2 \log N)$.

\subsection{Point spread function model}
Given the enormous computational complexity of the spatially variant 
convolution, we propose to use a neural network to emulate this process.
To this end, we compute the PSF as the autocorrelation of the generalized
pupil function $P$ \citep[e.g.,][]{vannoort05}, calculated here with
the aid of the inverse Fourier transform $\mathcal{F}^{-1}$:
\begin{equation}
  S(\mathbf{r}) = |\mathcal{F}^{-1}(P(\mathbf{v}))|^2.
  \label{eq:psf_autocorrelation}
\end{equation}
The generalized pupil function is given by the following general expression:
\begin{equation}
P(\mathbf{v})=A(\mathbf{v})e^{i \varphi(\mathbf{v})},
\label{eq:pupil_func}
\end{equation}
where $A(\mathbf{v})$ represents the amplitude of the pupil (we assume it fixed and known), and takes
into account the size of the primary mirror, the shadow of the secondary mirror 
or any existing spider, $\varphi(\mathbf{v})$ is the phase of the wavefront,
and $\mathbf{v}$ is the coordinate on the pupil plane. The imaginary unit number is denoted as $i$, as
usual. We choose to parameterize the phase of the wavefronts $\varphi(\mathbf{v})$ through 
suitable basis functions. This is a very efficient way of regularizing any 
blind deconvolution problem since PSFs are, by construction, sparsely parameterized
and are also strictly positive. In this study we use the
Karhunen-Lo\`eve \citep[KL;][]{karhunen47,loeve55} basis \cite[see also,][]{vannoort05}. {This
basis is constructed as a rotation of the Zernike polynomials \citep{1934Phy.....1..689Z}, with coefficients
found by diagonalizing the covariance matrix of the Kolmogorov turbulence \citep[see][]{roddier90}. Therefore:
\begin{equation}
\varphi(\mathbf{v})=\sum_{l=1}^{M}\alpha_{l}{\rm KL}_l(\mathbf{v}),
\label{eq:wavefront}
\end{equation}
where $M$ is the number of basis functions and $\alpha_{l}$ is the 
coefficient associated with
the basis function $\mathrm{KL}_l$. 

\subsection{The deep learning model}
Our proposal for SVCE is to use a conditional U-Net (shown in 
graphical representation in Fig. \ref{fig:unet}), a 
variation of the very succesful U-Net model of \cite{2015arXiv150504597R}
widely used in image-to-image problems. This model will produce
a discretized image, $i$, the result of convolving the object intensity, $o$,
with a spatially-variant PSF characterized by per-pixel KL coefficients, $\alpha$:
\begin{equation}
  i = f(x=\{o, \alpha\}),
  \label{eq:image_neural}
\end{equation}
where $f$ is the deep neural model described in the following. When discretized, 
$i$ and $o$ are 4D tensor of size $(B,1,X,Y)$, where $B$ is the batch size,
$X$ and $Y$ are the spatial sizes of the image and 1 is the number of channels. Similarly,
$\alpha$ is a 4D tensor of size $(B,M,X,Y)$, where $M$ is the number of KL coefficients.
Both $o$ and $\alpha$ are concatenated along the channel dimension to produce $x$, a 4D tensor
of size $(B,M+1,X,Y)$, and used as input to the neural network.
The graphical representation of Fig. \ref{fig:unet} clearly
shows two differentiated parts: an encoder that reduces the size of the images while
increasing the number of channels and a decoder that does the opposite operation. Some
information from the encoder is allowed to be passed to the decoder through
skip connections, which greatly improves the performance of the model. 
For clarity, we detail the operations carried out in each layer in the following.

\subsubsection{Encoder}
\label{sec:encoder}
The input tensor ($i$) is first passed through the following two sets of operations:
\begin{eqnarray}
  \hat p_1 &=& \mathrm{ReLU}(\mathrm{BN}(\mathrm{Conv}_3(C,x))), \nonumber \\
  p_1 &=& \mathrm{ReLU}(\mathrm{BN}(\mathrm{Conv}_3(C,\hat p_1))),
  \label{eq:doubleconv}
\end{eqnarray}
where $\mathrm{Conv}_3(C,x)$ is a 2D convolutional layer with $C=64$ kernels 
(we did not explore other values of $C$) of size $3 \times 3$ (the size of the kernel
is labeled with the subindex), followed by
batch normalization \citep[BN;][]{ioffe_batchnormalization15} and a rectified 
linear unit activation ($\mathrm{ReLU}(x)=\mathrm{max}(0,x)$). The
resulting tensor is then halved in size by using a $2 \times 2$ max-pooling
operation. This interchange of spatial information into 
channel information in the low resolution features $p_j$
efficiently exploits spatial correlation. After every max-pooling,
the resulting tensor is passed through the following
two operations:
\begin{eqnarray}
  \hat p_{j+1} &=& \mathrm{ReLU}(\mathrm{BN}(\mathrm{Conv}_3(2^{j+1}C,p_j))), \nonumber \\
  p_{j+1} &=& \mathrm{ReLU}(\mathrm{BN}(\mathrm{Conv}_3(2^{j+1}C, \gamma \hat p_{j+1} + \beta))),
\end{eqnarray}
where $j=1,\ldots,4$. These expressions are similar to those 
of Eq. (\ref{eq:doubleconv}), but with two 
scalars $\gamma$ and $\beta$ that are used for conditioning the
network on the specific instrument. The values of $\gamma$ and
$\beta$ are shared for all layers in the encoder.
This approach for conditioning is an extremely simplified version of the 
feature-wise linear modulation \citep[FiLM;][]{film17}, and the details
are explained in the following paragraph.
Additionally, the
number of channels is doubled with respect to the previous level.
The number of channels in the lowest resolution features is $2^4C=1024$,
and the spatial resolution of the input has been decreased by
a factor of $2^4=16$. This constitutes the bottleneck of the U-Net model and
turns out to be crucial for emulating very extended PSFs resulting from
very deformed wavefronts.

\begin{table*}[t]
  \centering
  \caption{Considered instrumental configurations.}
  \label{tab:instruments}
  \begin{tabular}{lcccccccc} 
    \hline
  Telescope & Instrument & Label & 1-hot & $D$ & $D_2$ & Pixel & $\lambda$ \\
  & & & vector & [cm] & [cm] & [arcsec] & [\AA] \\
    \hline
  SST & CRISP & CRISP$_{6302}$ & (1,0,0,0) & 100.0 & 0 & 0.059 & 6302 \\
  SST & CRISP & CRISP$_{8542}$ & (0,1,0,0) & 100.0 & 0 & 0.059 & 8542 \\
  SST & CHROMIS & CHROMIS$_{3934}$ & (0,0,1,0) & 100.0 & 0 & 0.038 & 3934 \\
  GREGOR & HiFI & HiFI$_{6563}$ & (0,0,0,1) & 144.0 & 40.0 & 0.050 & 6563 \\
    \hline
  \end{tabular}
  \end{table*}

We want this model to be as general as possible, so that it can be 
applied to telescopes of arbitrary diameter, any wavelength of
choice and any pixel size in the camera. This would require to 
condition the U-Net on a vector that encodes all this information. Training
such model would require to generate a suitable training set that 
covers all possible combinations. Since this is not realistic, and indeed 
is never the case in reality, we prefer to use a rather pragmatic
approach for conditioning. Given that only a few combinations of telescopes, 
cameras and wavelength are used routinely, we label each combination 
with an index $i=1,\ldots,I$, where $I$ is the total number of available 
combinations. We consider in this work the Swedish Solar Telescope (SST) with the
CRisp Imaging Spectropolarimeter \citep[CRISP;][]{2008ApJ...689L..69S}
and the CHROMospheric Imaging Spectrometer \citep[CHROMIS;][]{2017psio.confE..85S}, together with the 
GREGOR telescope with the High-resolution Fast Imager \citep[HiFI;][]{hifi2023}. The model can be easily extended
to other telescopes and instruments.
We transform each index into a one-hot vector\footnote{A vector
of length $I$, with a 1 on the component $i$ and zero elsewhere.} of 
length $I$.
These one-hot encodings are then passed through a small fully connected network (FCN) 
(displayed in pink in Fig. \ref{fig:unet}) 
with 2 hidden layers and 64 neurons in each layer. The 
output of this FCN produces the two scalars $\gamma$ and $\beta$.
We have not explored more complex conditioning schemes (e.g., vector conditioning
instead of scalar), but we argue that this simple global approach is sufficient for our purposes,
at least from what emerges from our results.

\subsubsection{Decoder}
In the decoder part of the U-Net, the features are doubled in 
size again by using linear upsampling until reaching
the original size. Skip connections propagating information from the
encoder at each scale are
added for the efficient propagation of gradients to all layers. This 
greatly accelerates the performance and training time of the U-Net model. 
After each upsampling and concatenation with the features of the 
same scale coming from the encoder, the following operations
are carried out:
\begin{eqnarray}
  \hat p_{j+1} &=& \mathrm{ReLU}(\mathrm{BN}(\mathrm{Conv}_3(2^{j+1}C,p_j))), \nonumber \\
  p_{j+1} &=& \mathrm{ReLU}(\mathrm{BN}(\mathrm{Conv}_3(2^{j+1}C, \hat p_{j+1}))),
\end{eqnarray}
where $j=4,\ldots,1$. Note that we do not condition the decoder on the
specific instrument configuration. We found it unnecessary but it can
be seamlessly added if needed. 
A final $1 \times 1$ convolutional layer produces the 
output $y$:
\begin{equation}  
  y = \mathrm{Conv}_1(1, p_{j+1}),
\end{equation}
Our specific model is fully convolutional, so that
the output and input spatial sizes are the same if  
the input tensor spatial sizes are multiple of $2^4$. Otherwise, the input 
and output can differ in a few pixels. The total number 
of trainable parameters of the neural model is 17.3M.


\subsection{Training}
The training of the model is inspired by the approach
followed by \cite{cornillere19}. The idea is to use a suitable
set of images with sufficient spatial variability and convolve
them with spatially invariant PSFs. Despite being trained with spatially
invariant PSFs, the fact that we supply the KL coefficients with spatial
structure produces a model that is able to learn spatially variant 
convolutions, as we demonstrate in Sec. \ref{sec:validation}.

Although the current status of 
solar magneto hydrodynamic simulations is very advanced, especially for
generating solar-like images, there are not 
enough simulations of quiet and active regions that are available for the community 
for a proper training. In an effort to overcome this difficulty, we decided to 
check whether the U-Net can be trained with images not specifically coming from solar
structures. From a conceptual point of view, the effect of a spatially 
variant convolution should be independent of the specific object.
We decided to use publicly available repositories of
images. One obvious possibility is \emph{ImageNet} \citep{imagenet_cvpr09}, that contains 
almost 15M high-resolution images, each one classified in a set of 1000 classes. We discarded this
option since the amount and sizes of the images is huge and makes downloading and training the
model very computationally demanding. A better option
for our purposes is \emph{Stable ImageNet-1K}\footnote{https://www.kaggle.com/datasets/vitaliykinakh/stable-imagenet1k},
a set of 512$\times$512 pixel images artificially generated with the Stable Diffusion 1.4 
generative model \citep{stable_diffusion}. The database
contains 100 images for each one of the 1000 classes of ImageNet. All images
are resized to 152$\times$152 pixels for training.

\begin{figure}
  \includegraphics[width=\columnwidth]{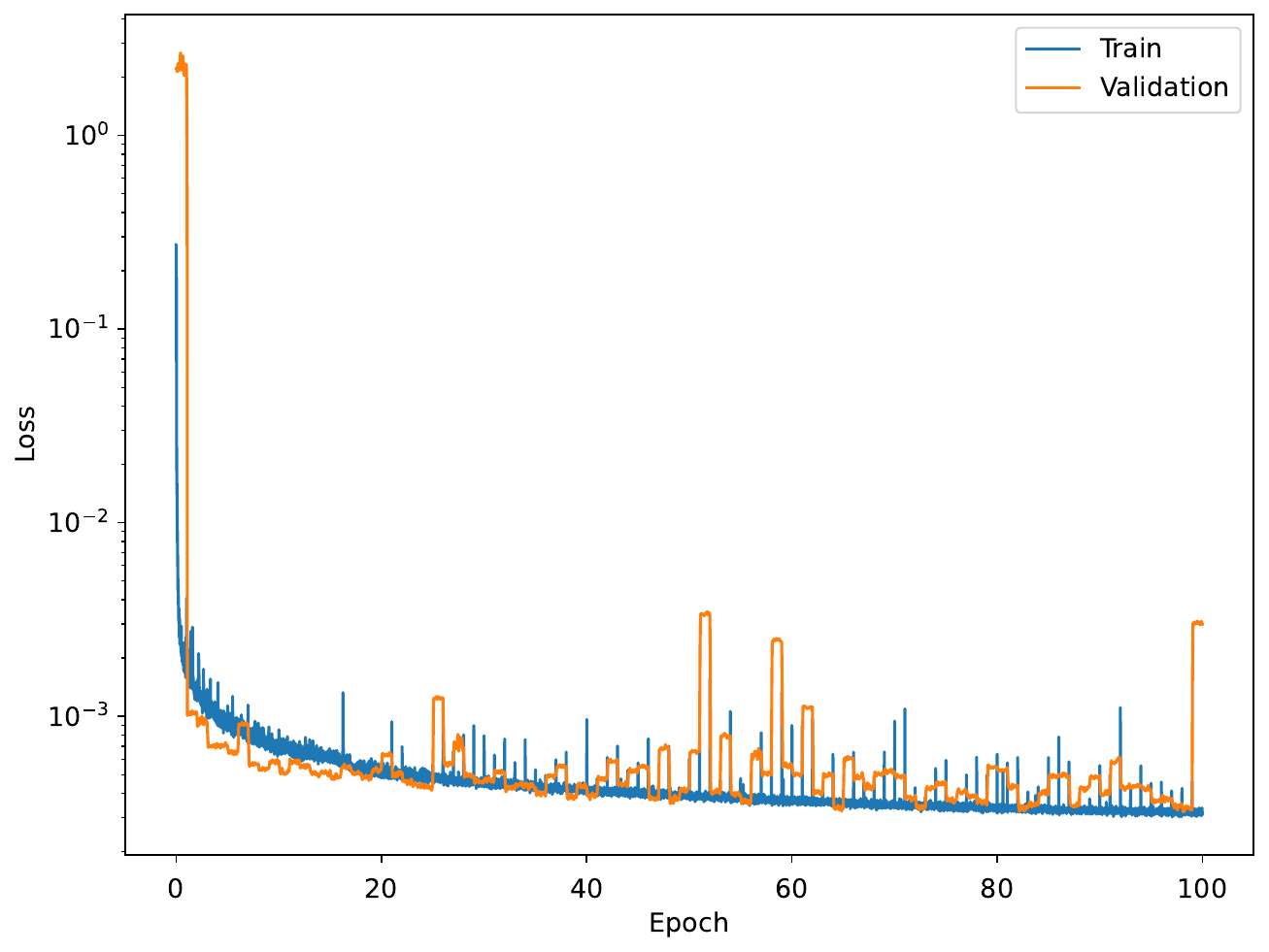}
  \caption{Training and validation losses evolution for all the
  epochs. The neural network does not show overtraining, although
  a saturation occurs after epoch $\sim$50.}
  \label{fig:loss}
\end{figure}

\begin{figure*}
  \includegraphics[width=\textwidth]{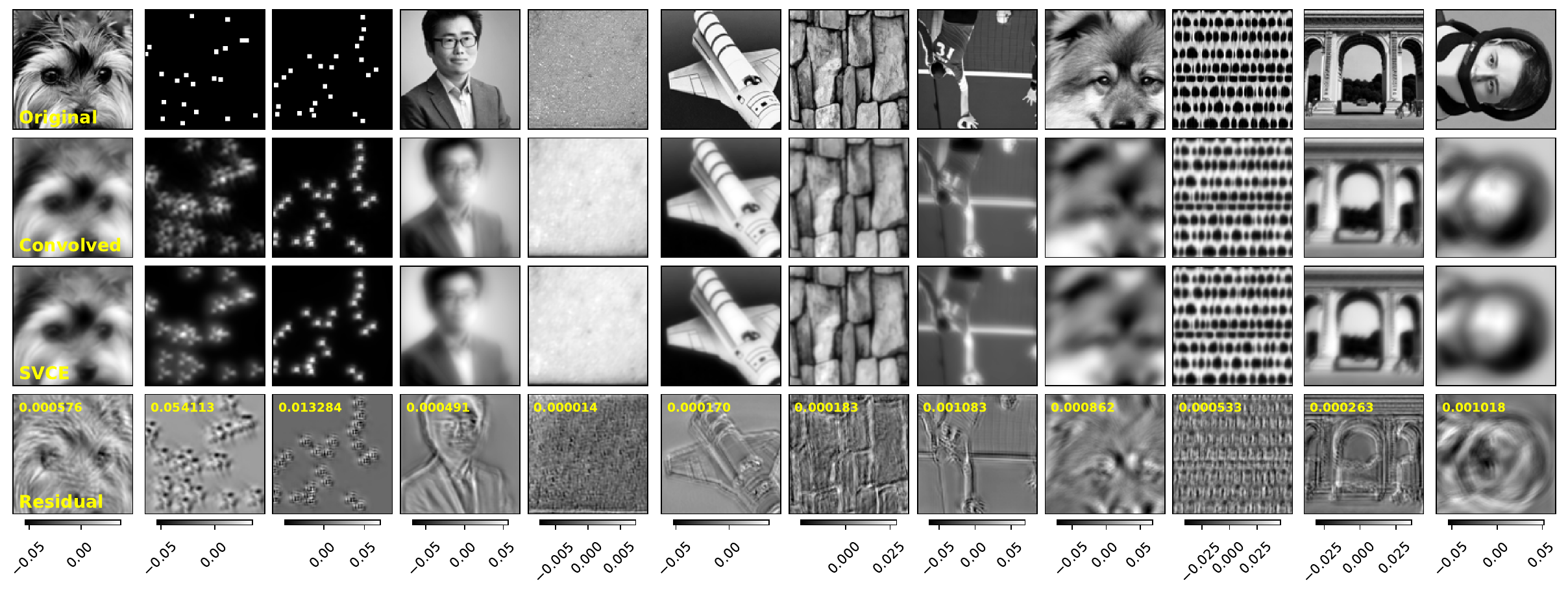}
  \caption{Samples from the training set, showing the capabilities of our 
  model. The upper row displays 12 original images from \emph{Stable ImageNet-1K} and
  synthetic image of point-like objects. The second row displays the target 
  convolved image, with a PSF that is compatible with Kolmogorov turbulence.
  The third rows shows the output of our model, with the fourth row displaying
  the residuals.}
  \label{fig:samples}
\end{figure*}

\begin{figure}
  \includegraphics[width=\columnwidth]{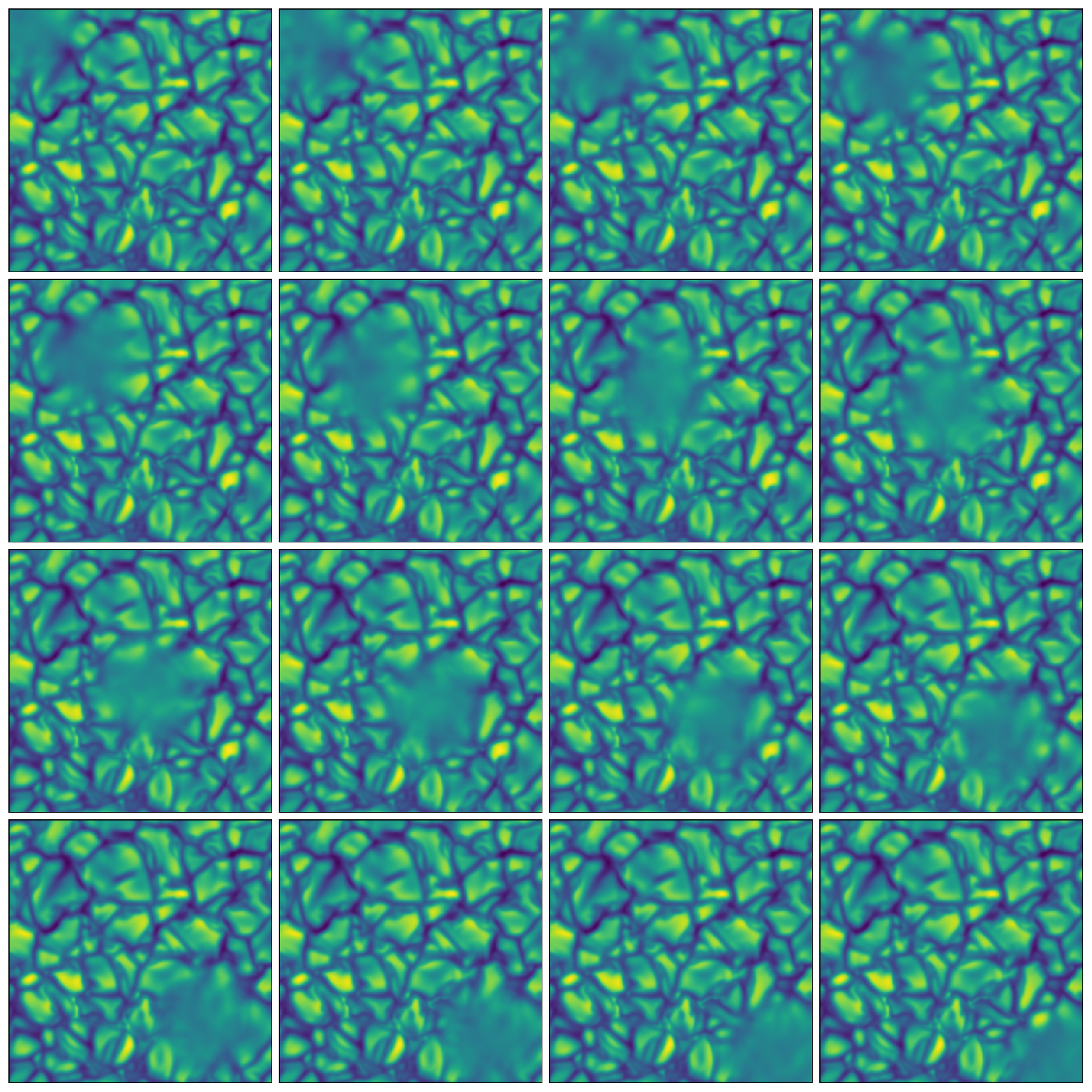}
  \caption{Spatially variant convolution with a defocus obtained with the SVCE. The defocus PSF
  is shifted along the diagonal of the image.}
  \label{fig:defocus}
\end{figure}

The PSFs are obtained from the wavefront coefficients using Eqs. (\ref{eq:psf_autocorrelation}), (\ref{eq:pupil_func}), and
(\ref{eq:wavefront}). The $\alpha$ coefficients are randomly sampled using the approach 
of \cite{roddier90}. The PSF depends on the telescope diameter $D$, central obscuration $D_2$, 
pixel size and wavelength. All considered combinations are shown 
in Tab.\ref{tab:instruments}. We consider random wavefronts with turbulence 
characterized by a Fried radius ($r_0$) in the range 3-15 cm. 
The variance of the KL modes are proportional to $(D/r_0)^{5/3}$, so that
smaller Fried radii lead to more deformed wavefronts, which produce
more complex and extended PSFs.
Every image from \emph{Stable ImageNet-1K} 
is then convolved with the computed PSF using the FFT, thanks to the assumption of a
spatially invariant PSF. Given that the images are non-periodic, we utilize
a Hanning window with a transition of 12 pixels in each border to apodize them so
that artifacts are avoided when using the FFT. Convolved images are then
cropped to 128$\times$128 to remove the apodized region. With this approach,
90k such images are used for training and 10k for validation. To increase the variability,
we include additional augmentations by randomly rotating (0$^\circ$, 90$^\circ$, 180$^\circ$ and 270$^\circ$)
and flipping horizontally and vertically the images. In our initial experiments using
only images from \emph{Stable ImageNet-1K}, we
realized that the model was not properly learning the specific shape of the PSF,
although the output of the model was very similar to the target. 
For this reason, and to force the model to properly learn the shape of the PSF, 
we substituted 20\% of the images of the training set by fields of randomly
located point-like objects in a black background. The convolution of these images with 
a PSF gives the shape of the PSF repeated several times in the image. After adding
them to the training set, we checked that this helped a lot in 
recovering a much better shape of the PSF, as shown below.

The training proceeds by using the output of the network, $y$ and
the target convolved image, $t$, and minimizing the following loss
with respect to the weights of the U-Net and those of the conditioning network:
\begin{equation}
  \mathcal{L} =  \left\| y-t \right\|^2 + 
  \lambda_\mathrm{grad} \left\| \nabla y - \nabla t \right\|^2 + 
  \lambda_\mathrm{flux} \left\| \langle y \rangle - \langle t \rangle \right\|^2.
\end{equation}
The first term is proportional to the mean squared error, that forces the per-pixel output
image to be as similar as possible to the target, measured with 
the $\ell_2$ norm. The second term forces the horizontal
and vertical gradients to be as similar as possible. We realized that adding
this term produced an improved quality for the convolutions, specially for 
the point-like objects. The gradients are implemented 
using Sobel operators. Finally, a third term forces the total
flux of the image to be preserved.
The hyperparameters $\lambda_\mathrm{grad}=10$ and
$\lambda_\mathrm{flux}=1$ produced good results. We leave for the
future a sensitivity analysis to the hyperparameters, which will require 
lots of computing time.

We trained for 100 epochs, ending up with the model that gives the
best validation loss. We used the Adam optimizer \citep{Kingma2014}
with a learning rate that starts at 3$\times 10^{-4}$ and is slowly
decayed to 6$\times 10^{-5}$ following a cosine rule \citep{DBLP:conf/iclr/LoshchilovH17}. 
The model
is trained on a single NVIDIA GeForce RTX 2080 Ti GPU using half precision, to
make a more efficient use of the video memory and tensor cores
onboard the GPU. The training is implemented and run using 
PyTorch 2.0.1 \citep{pytorch19}.

\begin{figure*}
  \centering
  \includegraphics[width=\textwidth]{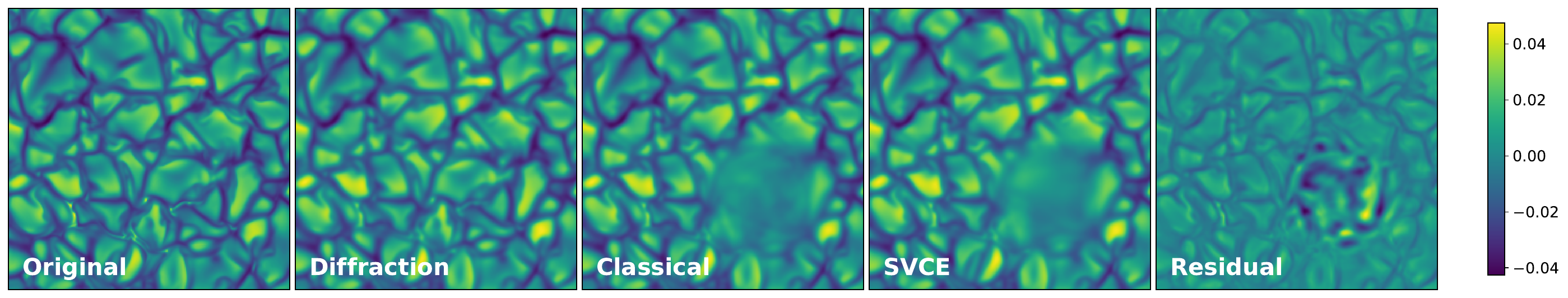}
  \caption{Original image from the \cite{stein12_a} simulation
  (first panel), together with the image after diffraction when
  observed with SST/CRISP at 8542 \AA\ (second panel). The third 
  panel displays the spatially variant convolution with a region of defocus
  computed using per-pixel convolution, with the fourth panel
  showing the results obtained with the neural SVCE. The last panel
  displays the residuals, with an NMSE of 1.6$\times$10$^{-4}$.}
  \label{fig:comparison}
\end{figure*}

Figure \ref{fig:loss} shows the training and validation losses during
training. The training proceeds nominally although the validation set 
clearly shows a saturation of the capabilities of the model around 
epoch 50. Along the training epochs, we find a rather variable validation loss, mainly 
produced by the strong penalization in the point-like objects
when the proper PSF is not produced in the output. Anyway, we only
see some hints of overfitting at the very end of the training. To overcome
it, we select the model producing the best validation loss as the optimal for our selection of 
hyperparameters.

\subsection{Validation}
\label{sec:validation}
Figure \ref{fig:samples} shows some examples of the behavior of the SVCE
on the validation set. The upper row shows either the original unperturbed images 
from \emph{Stable ImageNet-1K} or the synthetic point-like images. Some images
appear rotated due to the augmentation process. The second and third rows show the target 
image and the output of the SVCE, respectively. Finally, the 
fourth row displays the scaled residual between the target and the output
of the model. Since images are normalized to the interval $[0,1]$,
the residual shows differences below 5\%. The labels indicate the
normalized mean squared error (NMSE), computed as:
\begin{equation}
  \mathrm{NMSE} = \frac{\left\| y-t \right\|^2}{\left\| t \right\|^2}.
\end{equation}
Visually, both the target and 
the output of the model look very similar. Differences are, however, more
conspicuous for the point-like images. Very complex PSFs with lots of speckles
are difficuly to be reproduced. We argue that improved models can be obtained by carrying
out a hyperparameters tuning phase as well as using solar synthetic
images during the training. Improvements to the U-Net model can also produce
better final convolved images. We are, though,
happy with the trained SVCE given its proof-of-concept character, 
but work along the lines described above can decisively impact positively on
the quality of the SVCE and the ensuing reconstructions.

\begin{figure}
  \includegraphics[width=\columnwidth]{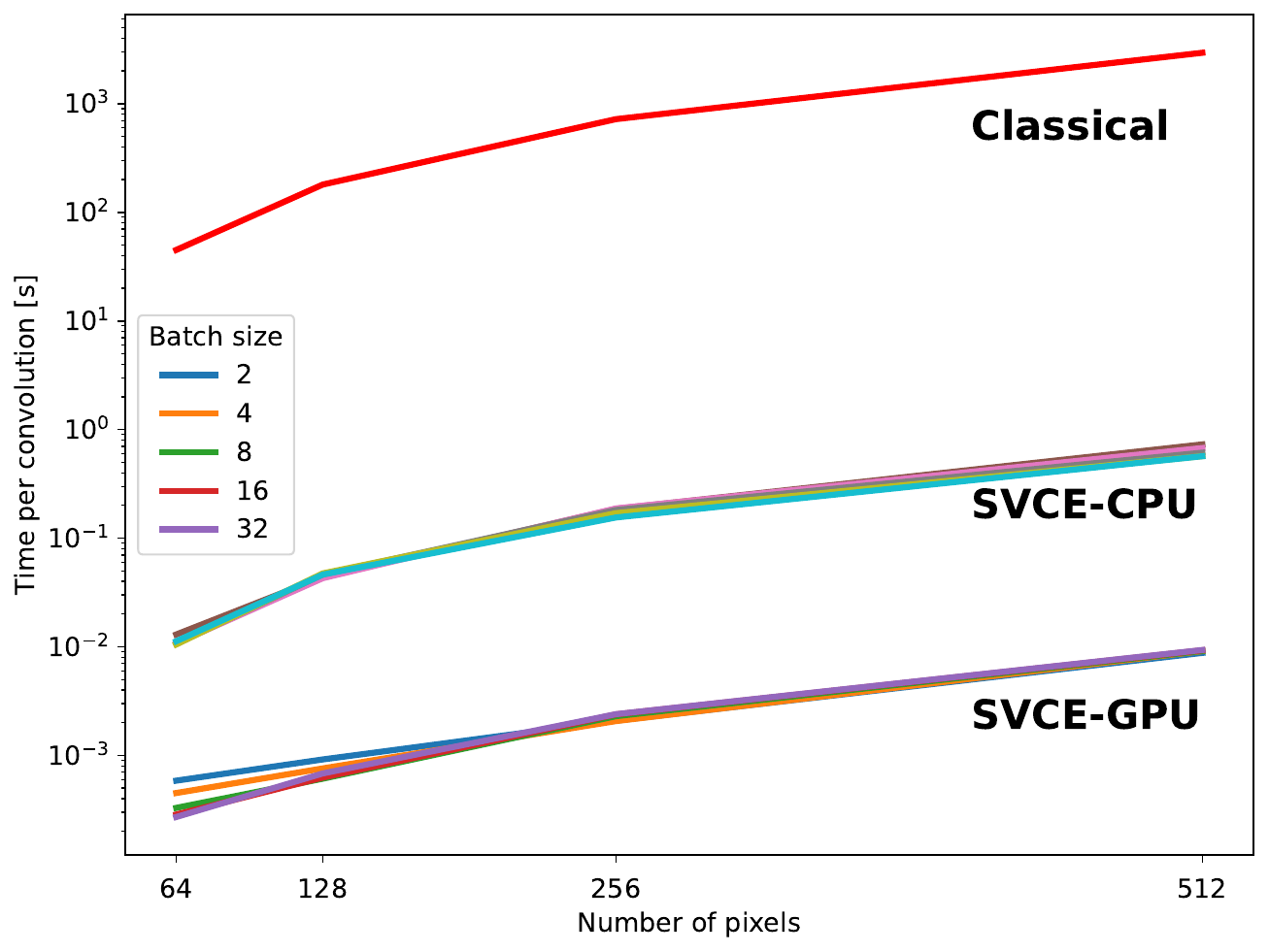}  
  \caption{Computing time per convolution as a function of the size of the image and
  the batch size for the SVCE, both in CPU and GPU. As comparison, we show results
  obtained for the classical method of computing the convolution of the image with
  a PSF for every pixel.}
  \label{fig:benchmark}
\end{figure}

Given that the neural model has been trained with synthetic images of the \emph{Stable ImageNet-1K}
database, it is important to test how it generalizes to other type of images. In particular, we
need to test whether the method does a good job in typical solar images. To this end,
we have used a continuum image from a snapshot of the quiet Sun hydrodynamical 
simulation of \cite{stein12_b} \citep[see also][]{stein12_a}. The pixel size of
this snapshot is 0.066 arcsec, which is close enough to the pixel size of the CRISP instrument.
Consequently, we have used the CRISP$_{8542}$ configuration for the SVCE.
Figure \ref{fig:defocus} shows the ability of the model to produce spatially 
variant convolutions in a subfield of 256$\times$256 pixels of the original
snapshot. We demonstrate how the neural SVCE is able to
mimick the effect of a defocus PSF that is gradually shifted along the diagonal
of the image. To this end, we simply set the KL coefficient for defocus equal
to a Gaussian centered on pixels along the diagonal with a standard
deviation such that it affects several granules.
The neural model is able to seamlessly reproduce this spatially
variant convolution running the neural model in evaluation mode. Given the
excellent paralellization capabilities of GPUs, if
enough GPU memory is available, all images in Fig. \ref{fig:defocus} can
be obtained in a single forward pass.

In order to test the reliability of the spatially variant convolution
of the SVCE, we have compared the results with the ones obtained
using classical techniques. To this end, we compute the convolution
of the original image with $256^2$ different PSFs, each one corresponding to the defocus found in 
every single pixel $(i,j)$. The construction of the final
image loops over all pixels in the output image, and the output is computed
by multiplying the PSF associated with the pixel (centered on the pixel) and the original image. 
This can be interpreted as an OLA method with a
1-pixel mosaicking window.
In our machine, the computation of the 256$^2$ convolutions takes $\sim 12$ minutes in CPU, which
is dominated by the computation of the PSF from the per-pixel wavefront coefficients.
The neural model takes $\sim 100$ ms in the same CPU, without any
effort in optimization, which is a speed-up of $\sim 3000$ times. A large part
of this gain is produced by not having to compute the PSF for every pixel from
the wavefront coefficients. Benchmarks are displayed in Fig. \ref{fig:benchmark}, where
the computing time per convolution is shown as a function of the size of the image and
the batch size. The SVCE is able to process images several orders of magnitude faster
than the classical method.

The comparison is shown in Fig. \ref{fig:comparison}. Our SVCE is
able to produce very similar results with a significant reduction in 
computing time. The neural SVCE does not require any apodization, so it 
works even at the borders of the image. For comparison, we show synthetic 
diffraction image for a telescope of 1m, together with the original image 
from the simulation. The SVCE is able to reproduce the diffraction
image away from the defocus point and a defocused image at the position
of the defocus. The maximum absolute residual is of the order of 4\% (images 
are normalized to 1), with a NMSE of 1.6$\times$10$^{-4}$.

\begin{figure}
  \includegraphics[width=\columnwidth]{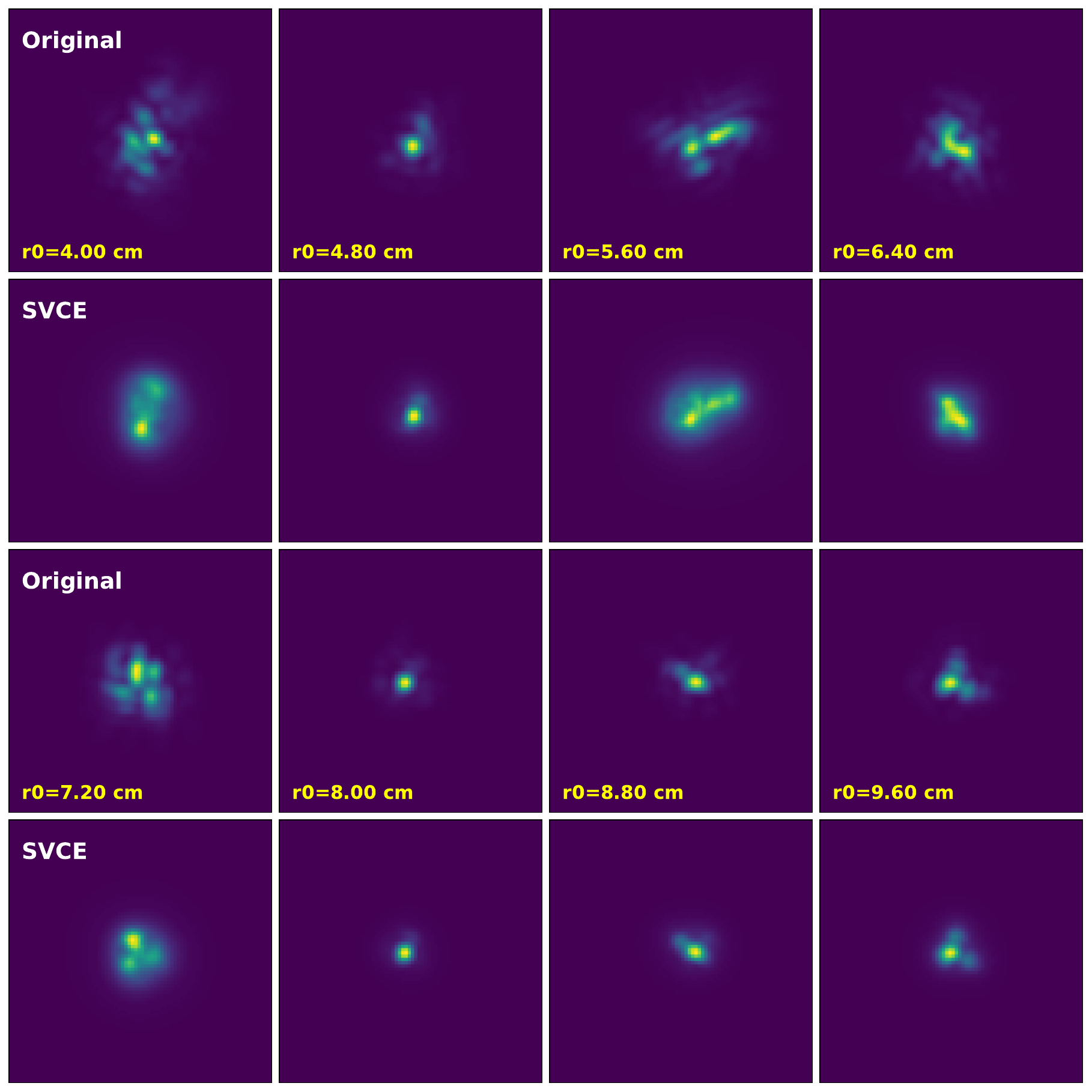}  
  \caption{Comparison of the exact point spread functions obtained for a specific
  realization of a Kolmogorov atmosphere with different 
  values of the Fried radius and the neural approximation. Although it captures
  the general behavior of the PSF, the neural model behaves worse for
  larger turbulence.}
  \label{fig:defocus_point}
\end{figure}

The effect of the SVCE on point-like objects is displayed
on Fig. \ref{fig:defocus_point}. We show the approximate PSF obtained with 
the neural model, and it is compared with the
correct PSF obtained from the KL modes. We show results for different realizations of Kolmogorov
turbulence with different Fried radii. The neural model is able
to nicely approximate the PSF for weak and medium turbulence, although 
its capabilities are degraded for very strong turbulence. This is demonstrated
in Fig. \ref{fig:nmse_r0}, where the NMSE is shown
to monotonically decrease for larger values of the Fried radius. 
This would require improvements on the training, as described above. Anyway, we 
argue that the general shape of the PSF is correctly recovered, which is enough for 
a good deconvolution of observations, especially when adaptive optics is used.

\section{Multi-object multi-frame spatially variant blind deconvolution}
The trained neural SVCE can be used as a replacement of the 
computationally intensive forward model
in multiframe blind deconvolution problems in cases
in which the PSF is spatially variant. Since the SVCE can deal
with per-pixel PSFs, one can get rid of the mosaicking process
and convolve large images much faster. In this section
we analyze its performance on standard optimization-based methods similar to the one proposed
by \cite{lofdahl_scharmer94} and \cite{vannoort05}. We show that the SVCE
produces a significant gain in computing time, allowing us to carry out image correction
in a full FOV in one go. In a set of reduced experiments we show that
it is possible to obtain reconstructions of good quality, almost comparable
to those obtained with the version of the MOMFBD code that we have used \citep{2021A&A...653A..68L}.

\subsection{Formulation of the problem}
Following the standard procedure, let us consider that we
simultaneously observe $K$ objects (e.g., monochromatic images of the
same region of the Sun at two different wavelenghts) and collect $L$ short exposure 
frames with a ground-based instrument of a stationary object outside Earth's atmosphere. 
All objects at the same frame are assumed to be affected by exactly the same wavefront.
In our case, given our scientific interests, our object is a small region on the surface of the Sun, but
the results can be applied to any other point-like or extended object. We note, in passing, 
that if this method is to be applied to point-like objects, better results would
be obtained by training the neural SVCE only with point-like objects. 

\begin{figure}  
  \includegraphics[width=\columnwidth]{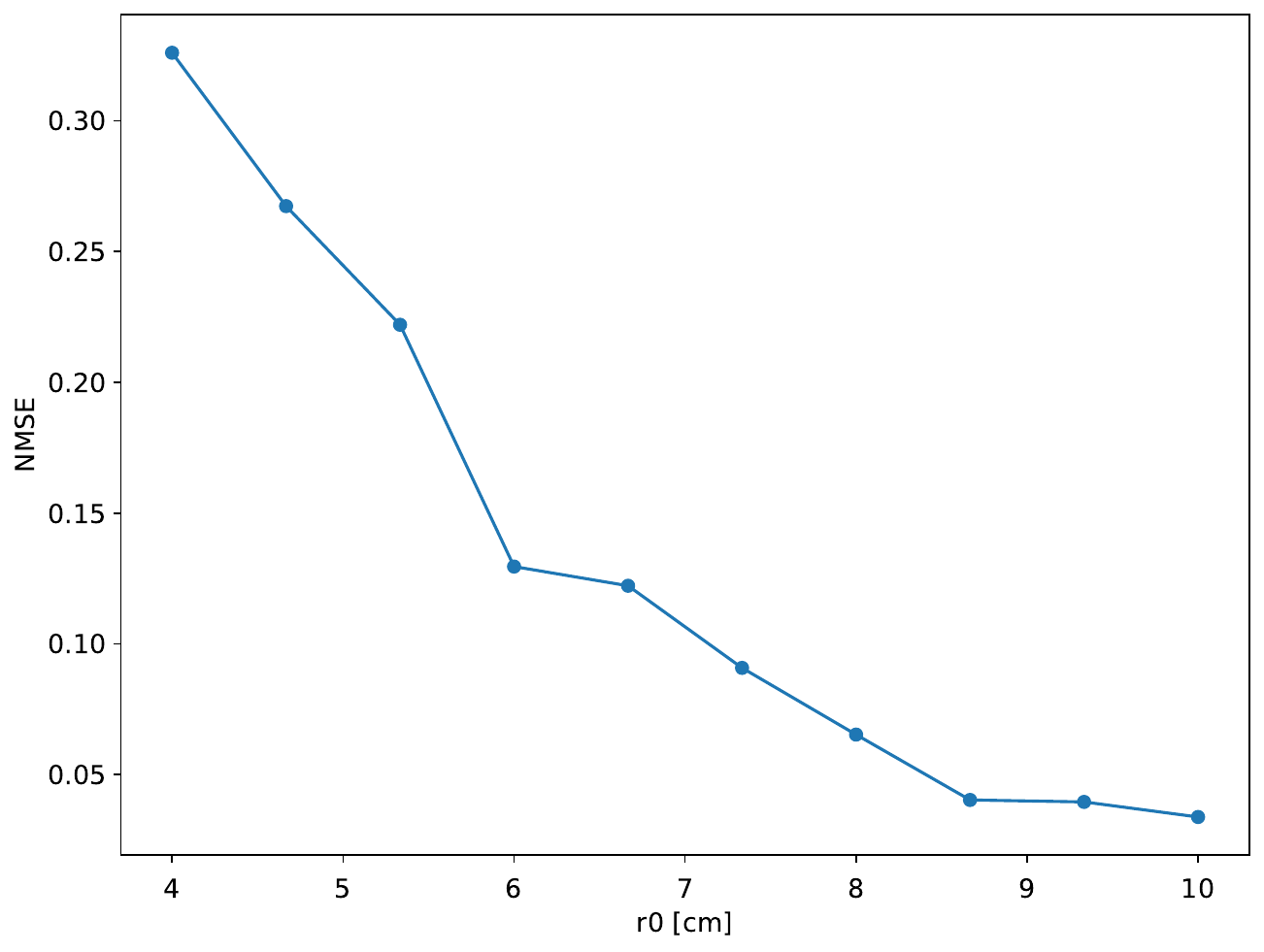}
  \caption{Comparison of the exact point spread functions obtained for a specific
  realization of a Kolmogorov atmosphere with different 
  values of the Fried radius and the neural approximation. Although it captures
  the general behavior of the PSF, the neural model behaves worse for
  larger turbulence.}
  \label{fig:nmse_r0}
\end{figure}

The exposure time
of each individual frame is short enough --of the order of a few milliseconds-- so 
that one can assume that the atmosphere
is ``frozen'', i.e., it does not evolve during the integration time. 
Additionally, we assume that the total duration of the burst is 
shorter than the solar evolution timescales, so that the
object can be safely assumed to be the same for all frames.
Let us represent the observed image of object $k$ at frame $l$ as $i_{kl}$.
As a consequence, using the neural SVCE of Eq. (\ref{eq:image_neural}), 
with its weights frozen, the observed image can be approximated as:
\begin{equation}
  i_{kl} = f(\{o_k ,\alpha_{l}\}) + n_{kl},
  \label{eq:gen_model}
\end{equation}
where $\alpha_{l}$ are the per-pixel wavefront coefficients (44 in our case because 
the SVCE was trained with that number of modes) for frame 
$l \in \{1 \dots L\}$ and $n_{kl}$ represents the noise term. The
SVCE is, consequently, used as a substitute of the forward problem, 
which would consist of mosaicking, convolving with per-patch PSFs and 
merging the patches together. Note that, since 
all objects observed simultaneously share the same atmospheric perturbations,
the PSF simply changes because of the potential
wavelength difference between the objects.

The spatially-variant multi-object multi-frame blind deconvolution problem (SV-MOMFBD)
consists of inferring the true objects $o_k$ and the
wavefront coefficients $\alpha_{l}$ from the observed images $i_{kl}$ using the SVCE
as a forward model. This is accomplished by maximizing the likelihood of the observed images. Under the 
assumption of Gaussian noise, the maximum-likelihood
solution can be obtained by minimizing the following loss function:
\begin{equation}
  L(o,\vect \alpha) = \sum_{k,l,x,y} \gamma_{kl} \left\| 
    i_{kl}-f(\{o_k,\alpha_{l}\} \right\|^2,
  \label{eq:loss}
\end{equation}
where the norms are calculated over all pixels for all frames and objects.
The weights $\gamma_{kl}$ are often set inversely proportional to the noise variance. For
simplicity, we have decided to select $\gamma_{kl}=1$ for all $k$ and $l$.
The assumption of Gaussian statistics is approximately valid in the solar case, in 
which the number of photons is large and the dynamic range of the illumination is small. If 
this is not the case but the number of photons is still large, one should take into account 
that the noise variance is proportional to the mean expected number of photons in each pixel. In low
illumination cases, one should resort to Poisson statistics.

\subsection{Regularization}
Even if the number of observed frames is large, the maximum likelihood
solution can lead to artifacts in the reconstruction. To minimize these
effects, one can apply early stopping techniques or use additional 
regularization techniques. 
We consider two different regularization techniques. The first one is
based on filtering the Fourier transform of the objects. In this case,
we substitute the object in Eq. (\ref{eq:loss}) by its filtered version:
\begin{equation}
  o^*_k = \mathcal{F}^{-1}\left(H \cdot \mathcal{F}(o_k)\right),
\end{equation}
where $H$ is a filter that removes the high frequencies of the object.
This idea was proposed by \cite{lofdahl_scharmer94} using a relatively complex and ad-hoc 
algorithm for computing $H$. Since that specific regularization cannot be 
easily applied to our case, we instead consider a simple Fourier filtering to reduce the presence
of frequencies above the diffraction limit. In this case, we have found good results
with a flat-top filter with a smooth transition to zero close to the diffraction
cutoff frequency. We build this filter as the convolution of a Gaussian and a rectangle, 
where $\nu$ is the spatial frequency in the Fourier plane in units of the 
cutoff frequency, $w$ defines the width of the filter and $n$ controls the smoothness of the transition:
\begin{equation}
  H(k) = \frac{1}{2} \left\{ \mathrm{erf} \left[ n \left(\nu + \frac{w}{2} \right)
  \right] - \mathrm{erf} \left[ n \left(\nu - \frac{w}{2} \right)
  \right] \right\}.
\end{equation}
Good filtering is found for $w=1.8$ (slightly less than the diffraction) and $n=20$.

We additionally explore Tikhonov regularization (i.e., 
$\ell_2$ regularization) for the
spatial gradients of the objects and the wavefront coefficients. This encourages
smooth solutions and helps reduce the artifacts. The final loss function is
given by:
\begin{equation}
  L(o,\vect \alpha) = \sum_{k,l}\gamma_{kl} \left\|
    i_{kl}-f(\{o^*_k,\alpha_{l})\right\|^2
  + \lambda_o \left\| \nabla o^*_k \right\|^2 + \lambda_\alpha \left\| \nabla \alpha_{l} \right\|^2,
  \label{eq:loss_regul}
\end{equation}
where $\lambda_o$ and $\lambda_\alpha$ are the regularization hyperparameters.
We point out that we will explore in the future other regularizations, such as total 
variation \citep[e.g.,][]{2023ApJ...951...59S}
or sparsity constraints \citep[e.g.,][]{2013A&A...552A.133S}, 
apart from investigating more elaborate differentiable filters in the Fourier plane.


\begin{figure}
  \centering
  \includegraphics[width=0.95\columnwidth]{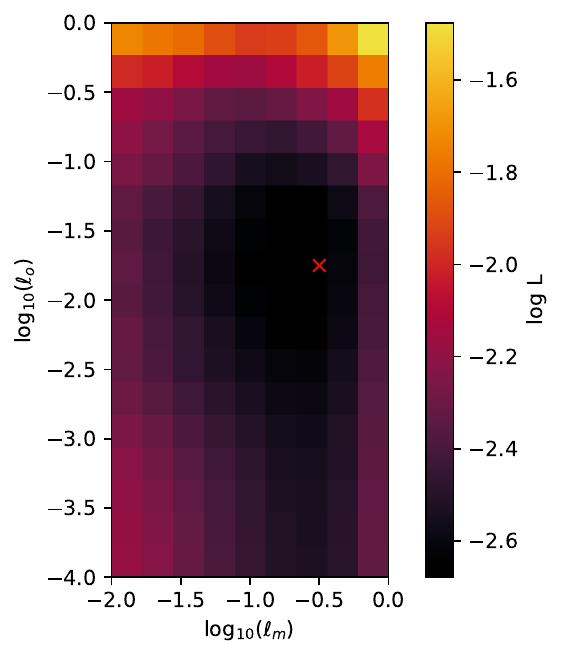}
  \caption{Hyperparameter search for the learning rates of the object and KL modes.}
  \label{fig:hyperparameters}
\end{figure}

\subsection{Optimization}
The loss function of Eq. (\ref{eq:loss_regul}) is simultaneously optimized with respect to 
the object and the wavefront coefficients using PyTorch, which
greatly facilitates the computation of the gradients using automatic differentiation. 
Note that the weights of the trained SVCE are frozen during this optimization. 
The optimization is carried out using the AdamW optimizer \citep{DBLP:conf/iclr/LoshchilovH19},
very similar to Adam. 
We have verified that using two different learning rates in the AdamW optimizer,
one for the objects, $\ell_o$, and one for the KL modes, $\ell_m$, gives excellent results. 
To find the optimal values
of these learning rates, we show in Fig. \ref{fig:hyperparameters} the
loss obtained after 20 epochs for the deconvolution of a $128 \times 128$
observation from the CRISP instrument at 8542 \AA. A clear optimal is
found for $\log_{10} \ell_o \approx -1.7$ and $\log_{10} \ell_m \approx -0.5$.
These values, marked with a red cross in Fig. \ref{fig:hyperparameters}, 
are used in all our subsequent results.

Our approach can infer the per-pixel wavefront coefficients. 
In this case, the number of unknowns is $X \times Y \times K$ for the objects and
$X \times Y \times M \times L$ for the wavefront coefficients. As a reminder, $X$ and $Y$ are the 
spatial dimensions of the images, $K$ is the number of objects, $M$ is the number of KL modes and
$L$ is the number of short exposure frames. Using typical values for the parameters, the number of
unknowns is of the order of 0.5M for the objects and 230M for the wavefront coefficients. In our experiments, 
the explicit smoothness regularization defined in Eq. (\ref{eq:loss_regul}) seems
to be enough. However, we introduce an extra regularization by reducing 
the number of unknown modes by only inferring the wavefront 
coefficients in a low-resolution grid, which is a factor $P$ smaller than the
original images. The wavefront coefficients in the whole image are computed 
using bilinear interpolation. In this case, the number of unknowns for the wavefront
coefficients is reduced to $X/P \times Y/P \times M \times L$. Using $P=4$, the number of unknowns
is reduced to 14.4M.

\begin{figure*}
  \centering
  \includegraphics[width=0.98\textwidth]{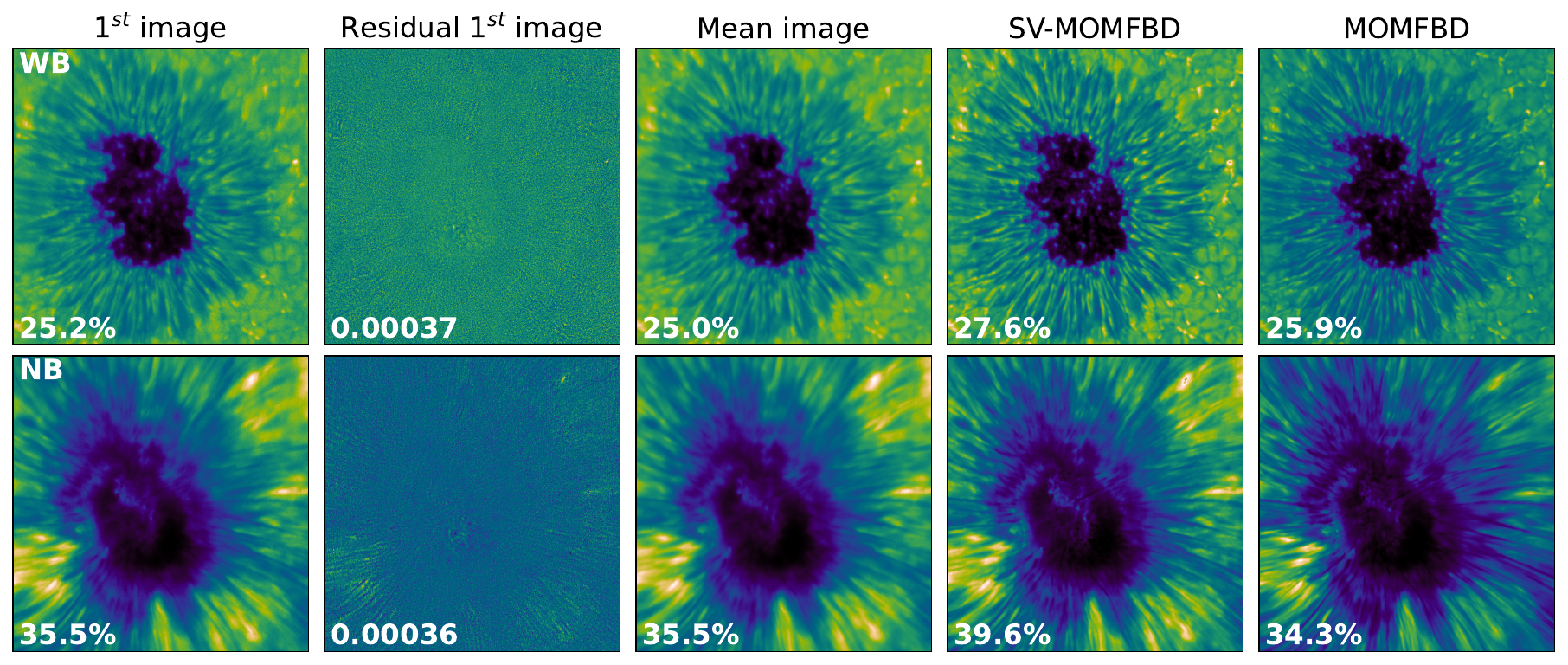}
  \includegraphics[width=0.98\textwidth]{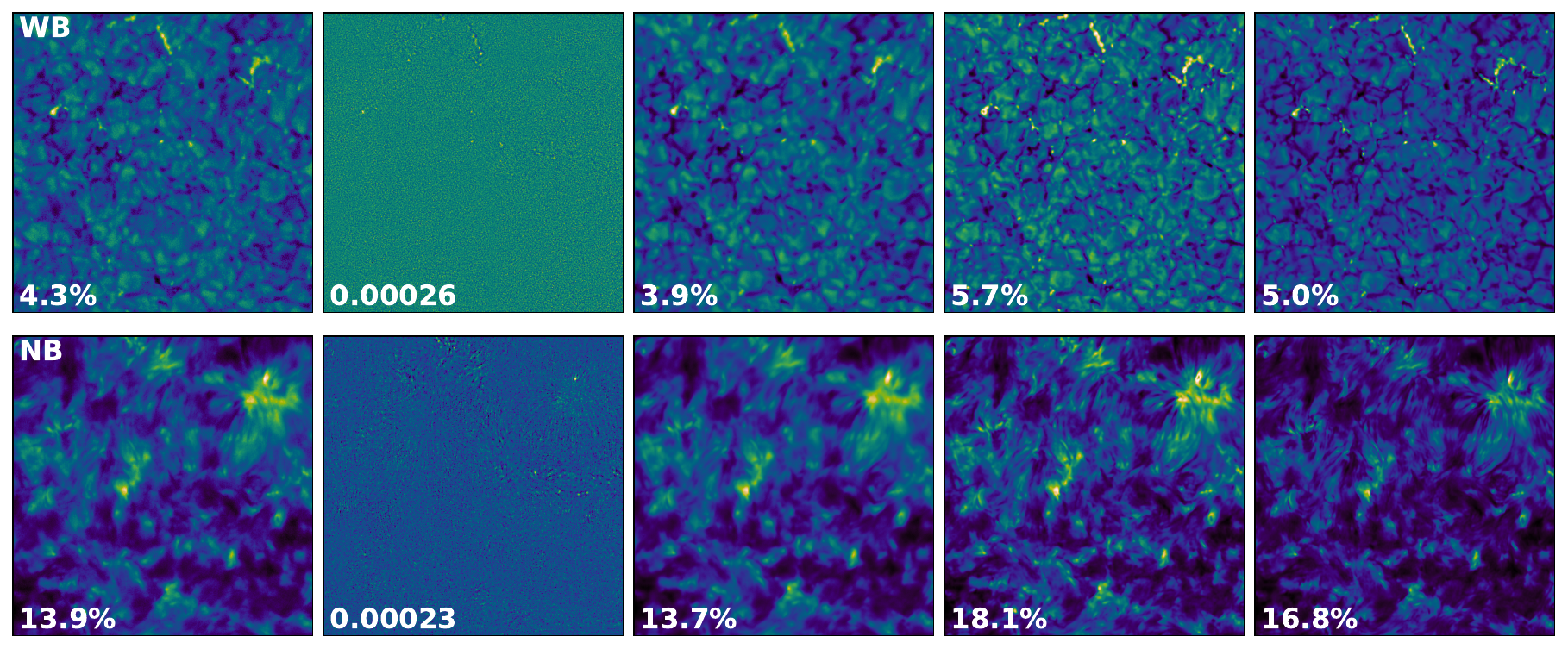}
  \includegraphics[width=0.45\textwidth]{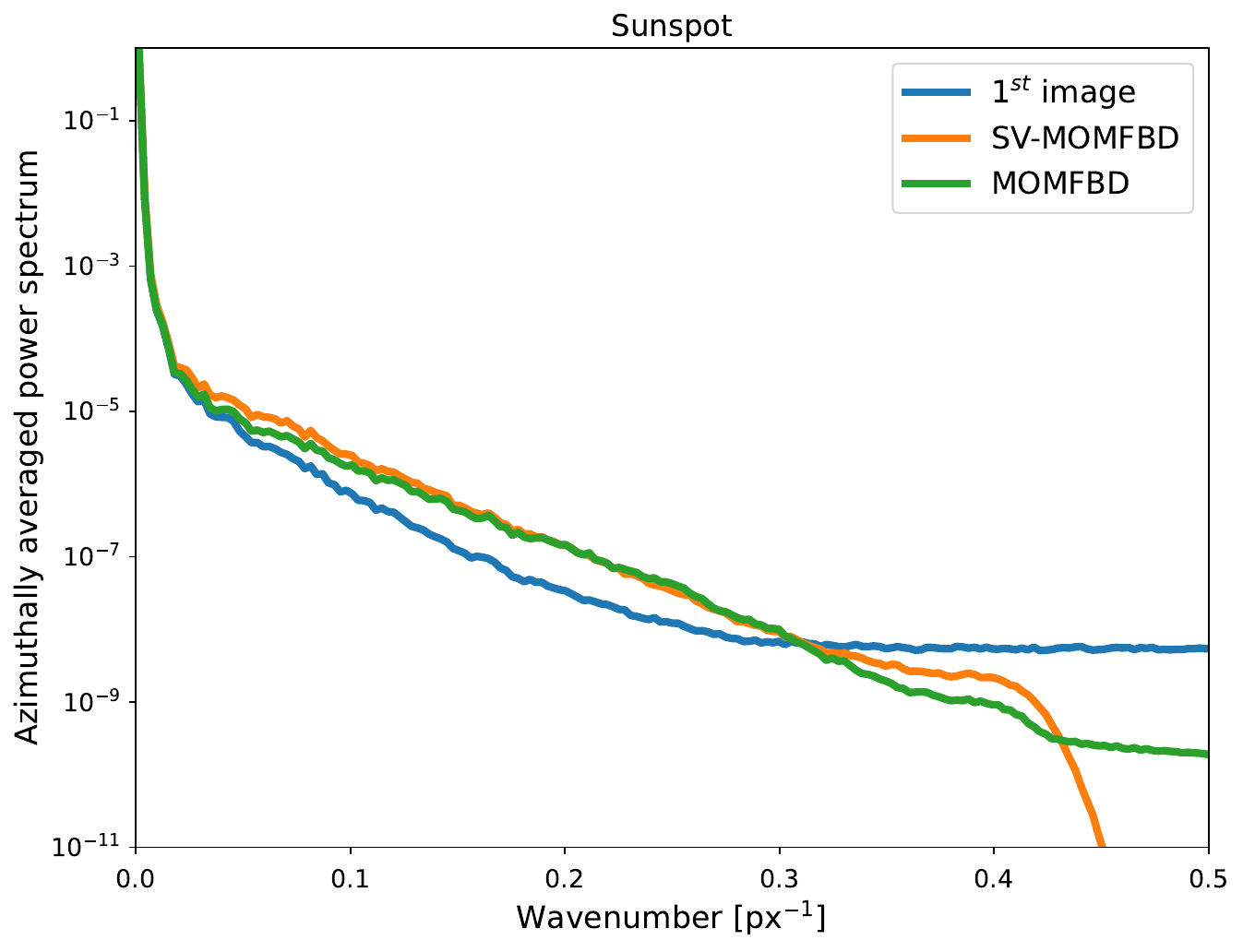}
  \includegraphics[width=0.45\textwidth]{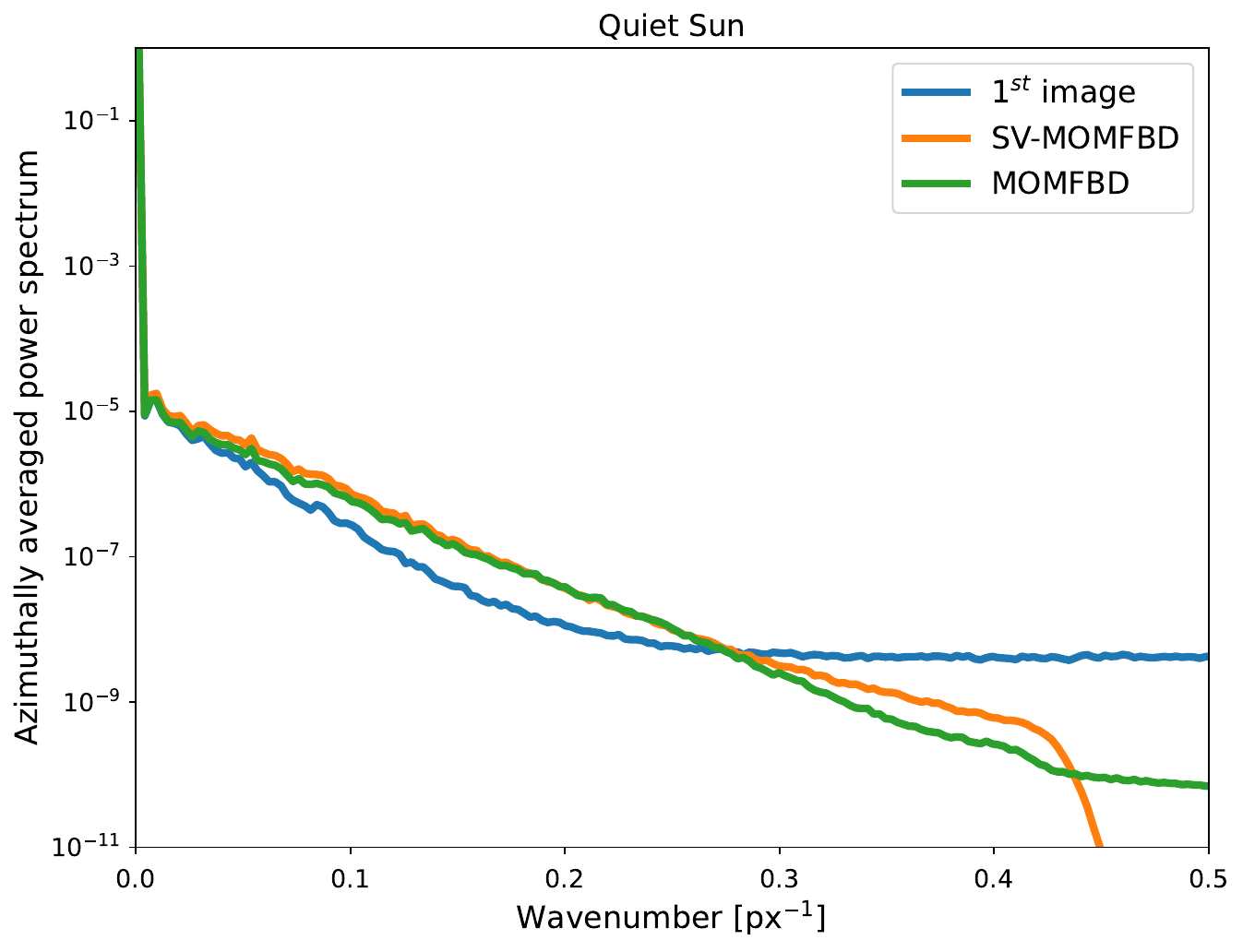}
  \caption{Image reconstruction results for CRISP for a sunspot (upper panels) and a
  quiet Sun (middle panels) observations. The upper row in each panel shows the results for
  the wide-band image, while the lower row shows the results for the narrow-band image.
  The first column shows the first frame in the burst. The second column shows the 
  residual between the first frame and the one obtained by convolving the
  inferred object with the inferred spatially variant PSF. The third column shows the average
  of all observed frames. The fourth and fifth columns display the reconstruction with 
  SV-MOMFBD and MOMFBD, respectively. The contrast of the images is quoted in all
  panels except for those of the second column, which show the NMSE.
  The lower panels show the azimuthally averaged spatial power spectra for the WB channels of 
  the sunspot and quiet Sun observations.}
  \label{fig:reconstruction_crisp}
\end{figure*}


The deconvolution can be carried out in the CPU or, ideally, using a GPU.
Given the potential limited amount of VRAM (Video Random Access Memory) of current
generation of GPUs, it is interesting to
apply several well-known techniques to scale our approach to very large images. The first 
method that we apply is doing all computations in half-precision (FP16), taking
advantage of mixed precision training \citep{DBLP:conf/iclr/MicikeviciusNAD18}. The amount of
VRAM used is practically halved with respect to using single precision (FP32), with an
additional signficant reduction in the computation time. In order to avoid 
numerical underflow, mixed precision training requires to scale the loss function to
avoid numerical underflow. The second method that we apply is gradient
checkpointing, that trades off VRAM usage for computation time during 
backpropagation. Backpropagation requires storing intermediate activations for 
computing gradients during the backward pass, which can be memory-intensive.
Instead of storing all activations, only a subset of them are stored. The remaining
ones are recomputed on-the-fly during the backward pass. 
With these techniques, a burst of 20 frames of size $512 \times 512$ with a single 
object can be deconvolved using only 8 GB of VRAM. Larger images or 
a larger number of objects can still be treated
either by using GPUs with more memory, using CPUs (which will make the 
optimization process typically an order of magnitude slower) or using 
mosaicking.

\section{Results}
\subsection{Reconstructions}
We demonstrate in the following the capabilities of our SVCE on
reconstructing observations of size $512 \times 512$. The quality of the reconstruction is shown in
Fig. \ref{fig:reconstruction_crisp} for two observations with CRISP at the SST. The upper panels correspond to the observation of active region AR24473 on 
27 Jul 2020, located at heliocentric coordinates $(-20",-416")$. The cadence of
the observations is 20 s and a burst of 12 frames is used for the reconstruction. 
No alignment or destretching is used in this case. The image stability
was good so that we assume that spatially-variant tip-tilts can be used during
the optimization to destretch the images. The lower
panels correspond to a quiet Sun observation at disk center, observed on 1 Aug 2019 with a
cadence of 31 s, also composed of a burst of 12 frames. The narrow-band (NB) images 
(lower rows in each panel) correspond to monochromatic images 
at 65 m\AA\ and 130 m\AA\ from the Ca \textsc{ii} 8542 \AA\ chromospheric spectral line. The
wide-band images (upper rows in each panel) correspond to the local continuum. We label each image
with the value of the contrast, defined as the standard deviation of the image
divided by the average. The first column displays $i_{k1}$,
the first frame of the burst, which gives an idea of the quality of the images with the
adaptive optics correction. The second column shows $i_{k1}-f(\{o_k,\alpha_1\})$, the
difference between the first frame and the one obtained by convolving the 
inferred object with the inferred spatially variant PSF. The square of this metric is
forced to be zero during the optimization. We quote the NMSE in this column.
The third column displays $\langle i_{kl} \rangle_{l=1\ldots L}$, the average of 
all frames in the burst. The noise is slightly reduced when compared with a
single frame, but the contrast is also reduced.
Since the observing conditions were good, the
average image is very similar to the first frame. The fourth column shows $o^*_k$, the
reconstructed image using the SV-MOMFBD (including filtering), which should be compared with the
result of the application of the MOMFBD code (last column).
We utilize $\lambda_o=0.1$ for the sunspot observations and $\lambda_o=0.13$ 
for the quiet Sun reconstructions. In both cases, $\lambda_\alpha=0.01$.
A small residual noise is still present in the reconstructions if
a smaller $\lambda_o$ is used. We optimize the loss function with the AdamW optimizer 
for 30 iterations. We infer the modes in a grid of $64 \times 64$, which
is a factor 8 smaller than the size of the images. The computing time per iteration is 0.75 s in an
NVIDIA RTX 4090 GPU. 

\begin{figure*}
  \includegraphics[width=\textwidth]{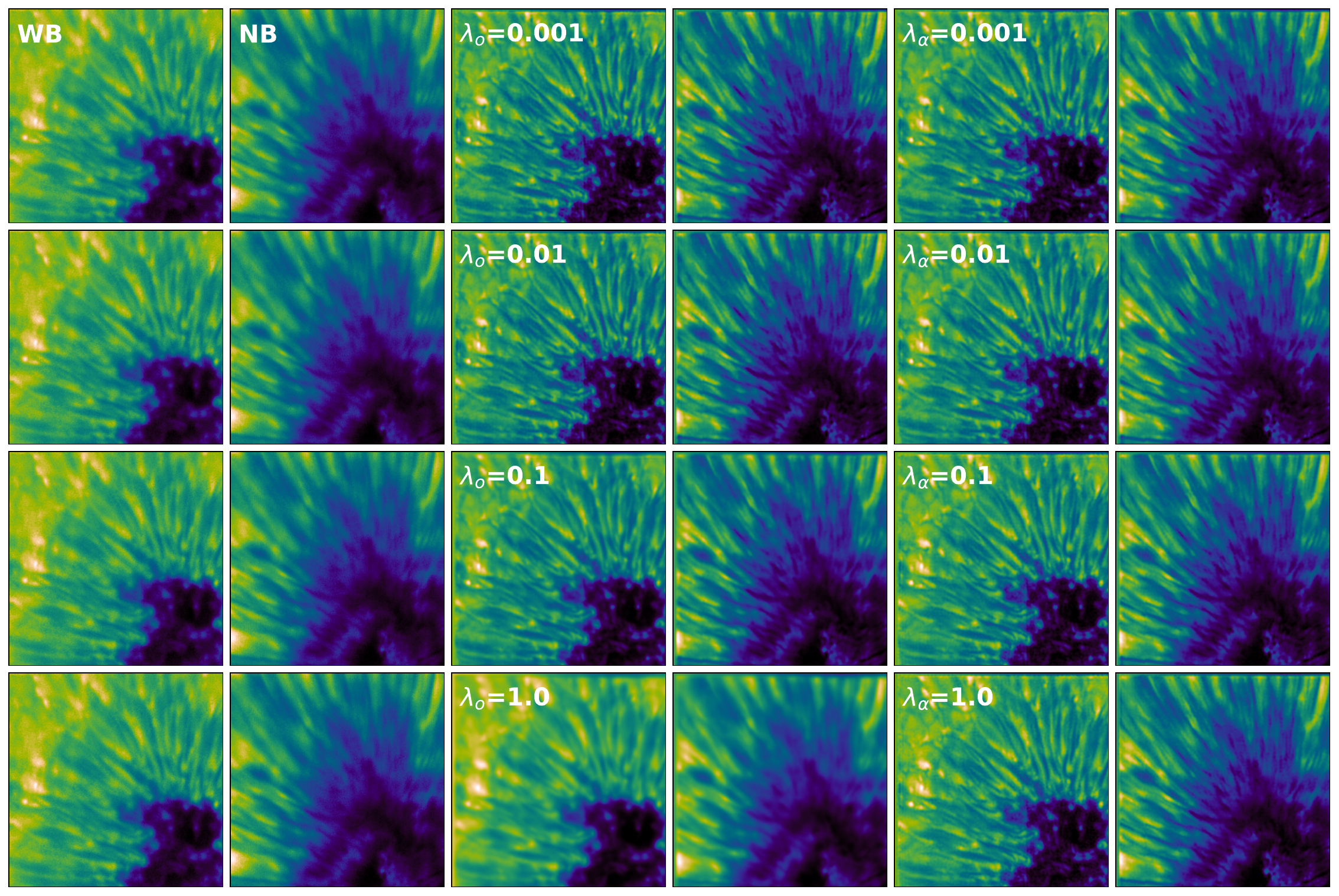}
  \caption{Impact of the hyperparameters for object and modes regularization. The first and second column 
  display the original WB and NB first frames fo the burst. The third and fourth columns show the
  reconstructed WB and NB images, respectively. Each row is labeled by the regularization
  parameter $\lambda_o$ (we use a fixed value of $\lambda_\alpha=0.01$). The fifth and sixth columsn show the reconstructions when 
  the regularization parameter $\lambda_\alpha$ is varied (we use a fixed value of $\lambda_o=0.01$).}
  \label{fig:regularization}
\end{figure*}

\begin{figure*}
  \includegraphics[width=\textwidth]{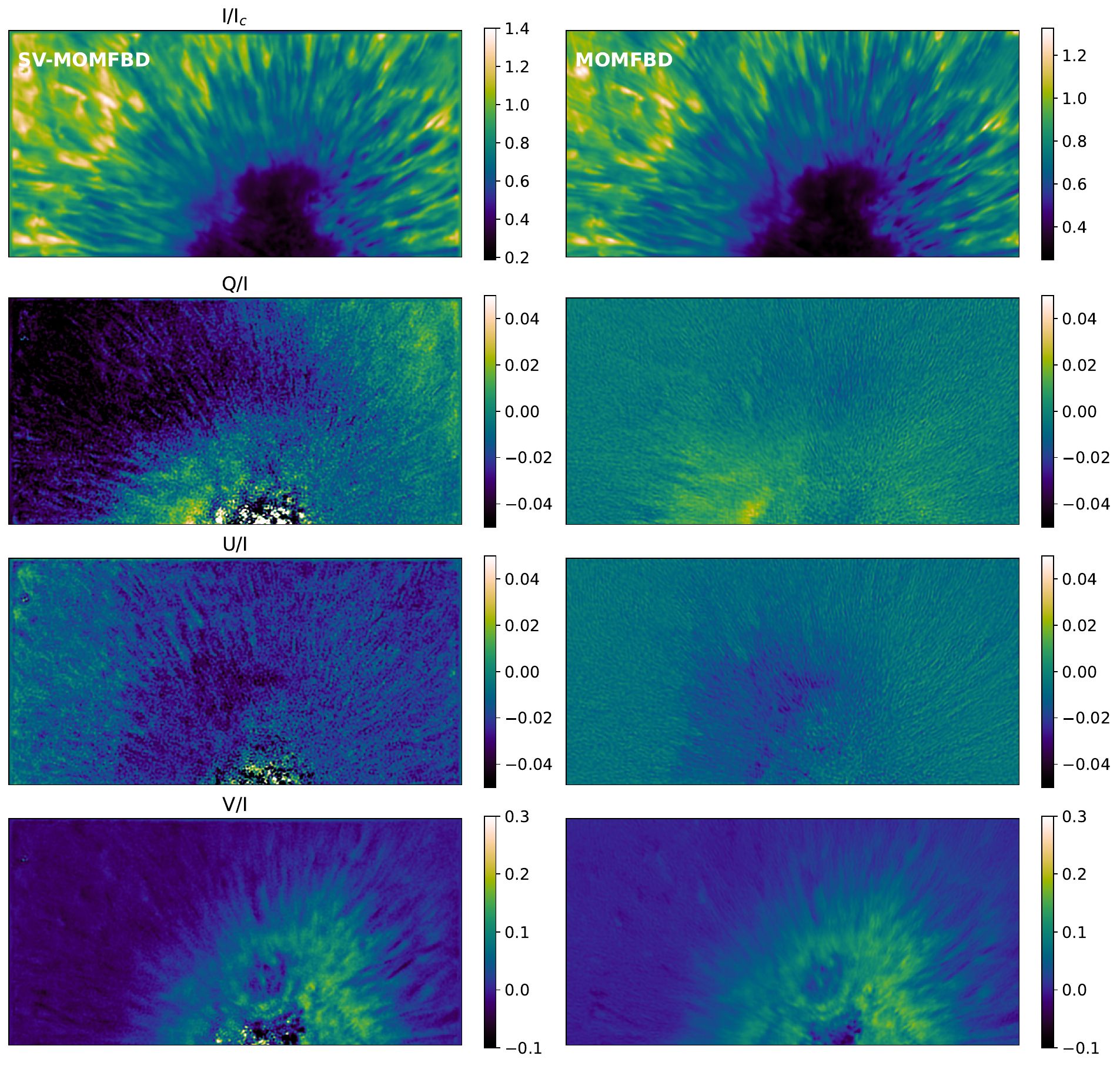}
  \caption{Monochromatic polarimetric images in the NB channel at 260 m\AA\ from
  the Ca \textsc{ii} 8542 \AA\ line center.}
  \label{fig:polarimetry}
\end{figure*}

Our reconstruction of the active region is able to correctly recover very fine details. The
comparison between SV-MOMFBD and MOMFBD is very good. Visually, the contrast is 
slightly larger in SV-MOMFBD than that found with MOMFBD, which is also quantitatively shown in the figures.
The dark cores in the penumbra found in the WB image are recovered visually similarly in both
approaches. The same happens for the umbral dots. Concerning the NB images, the contrast is 
also increased. Despite the similarities, it is true that the SV-MOMFBD reconstruction
has some issues. There are some bright regions that are recovered brighter with SV-MOMFBD
than with MOMFBD, especially in the NB channel. These regions are indeed present in the individual frames, but 
they seem to be absent from the MOMFBD reconstruction. Additionally,
some small artifacts in the form of rings seem to appear. The dark cores in the penumbra
in the NB channel look less straight in SV-MOMFBD than in MOMFBD. The original individual frames 
seem to contain this feature, which is somehow not absent in MOMFBD. We have not been able 
to improve them with hyperparameter tuning, so it is clear that more exploration (regularization,
improved SVCE, \ldots) is needed 
in the near future. Concerning the
dark filamentary structure, both reconstructions also produce similar results.
The quiet Sun reconstructions are also of very good quality, with contrasts that are slightly larger
when using the SV-MOMFBD. Again, the NB images show brighter regions that are not recovered
with MOMFBD, although these regions are present in the individual frames. Apart from that, we
obtain structures of very similar spatial scales than those found with MOMFBD.

\begin{figure*}
  \centering
  \includegraphics[width=0.49\textwidth]{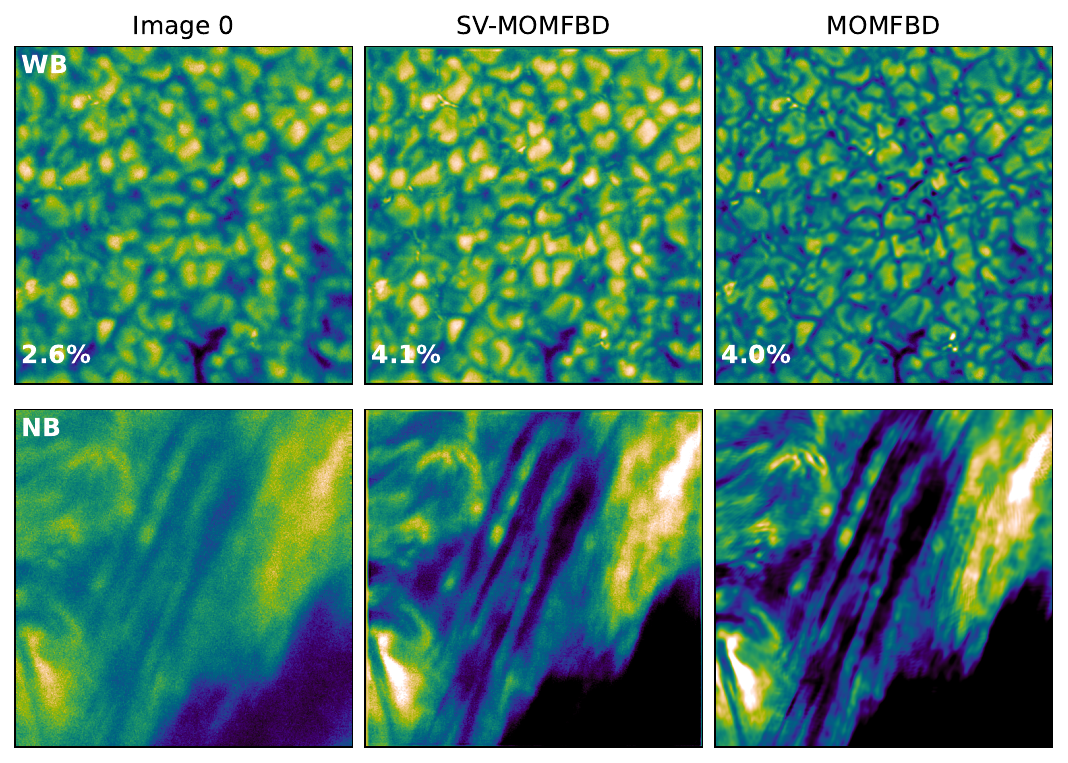}
  \includegraphics[width=0.49\textwidth]{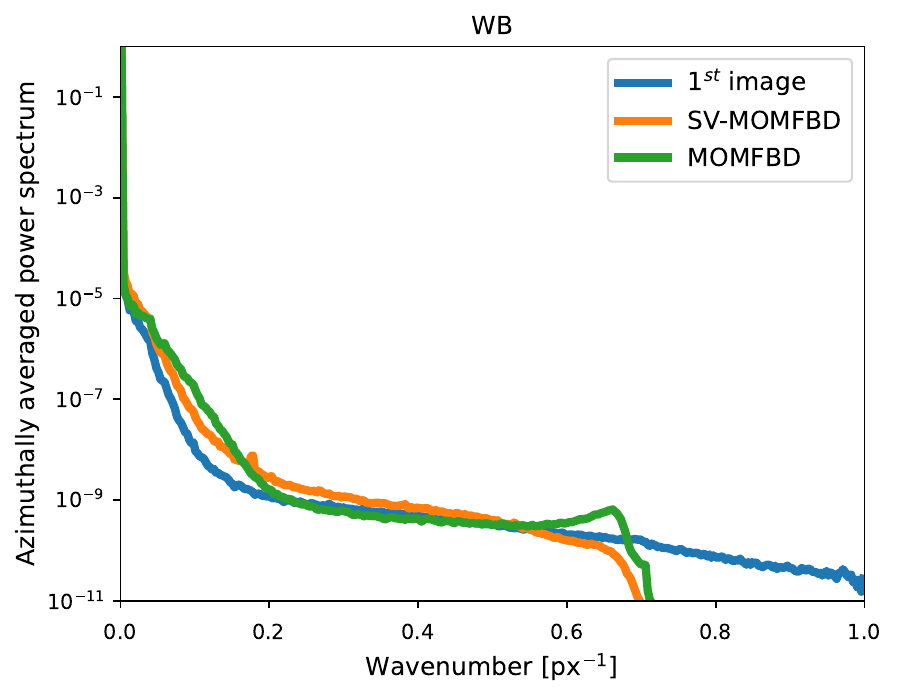}
  \caption{Example of reconstruction of HiFI data (left panel). The first column
  displays the first images of the burst in the WB and NB channels. The second and 
  third columns show the reconstructed WB and NB images with our SV-MOMFBD and MOMFBD, respectively.
  The right panel shows the azimuthally averaged spatial power spectra for the WB channels.}
  \label{fig:hifi}
\end{figure*}

For a more quantitative comparison, we show in the lower panel of Fig. \ref{fig:reconstruction_crisp} the azimuthally averaged
spatial power spectra for the WB channels for the sunspot and quiet Sun observations. The blue 
curves correspond to individual frames, with the plateau at large wavenumbers produced by the
noise. The green and orange curves correspond to the MOMFBD and SV-MOMFBD reconstructions, respectively.
The increase in spatial power between 0.02 px$^{-1}$ and 0.3 px$^{-1}$ is relevant
and similar in both reconstructions. This corresponds to the recovery of structures 
of size between 0.2" and 3", that have been destroyed by the seeing. The power spectrum closely
follows a power law, strongly reducing the noise for spatial information above 0.3 px$^{-1}$ when
compared to individual frames.
The noise reduction abilities of our SV-MOMFBD is very similar to that of MOMFBD. The tiny 
differences can be associated with the details of the Fourier filtering implemented in both codes. We note
a rather abrupt fall of the power above the diffraction limit in SV-MOMFBD. This fall is also visible
in MOMFBD, but then the power keeps an almost constant value. This high frequency residual
is produced by the stitching of the patches.

The effect of the Tikhonov regularization on the reconstruction is shown in Fig. \ref{fig:regularization}.
The first and second columns display the first frames fo the burst for the WB and NB channels,
respectively. The impact
of the regularization parameter $\lambda_o$ (while keeping $\lambda_\alpha=0.01$)
is shown in the third and fourth columns for the
WB and NB images, respectively. We note that the spatial frequencies of the reconstruction for $\lambda_o=1$
are severaly damped, with a significant loss of contrast. However, the reconstruction for 
$\lambda_o=10^{-3}$ is more affected by noise. This noise, which has spatial correlation, 
can trick the eye to produce excessively crispy images. We conclude that a value in the
range $\lambda_o=[0.01, 0.1]$ seems appropriate for our purposes. 
Meanwhile, the fifth and sixth columns show the effect of changing the regularization
parameter $\lambda_\alpha$ (while keeping $\lambda_o=0.01$). We find that the reconstruction
is almost insensitive to the specific value of $\lambda_\alpha$.
Anyway, the $\lambda_o$ and $\lambda_\alpha$ hyperparameters
should be taylored for each specific observation and need to be tuned
by the user.

Maintaining the photometry is important to keep the polarimetric information
after image reconstruction. We show in Fig. \ref{fig:polarimetry}
the monochromatic Stokes images at 260 m\AA\ from the line center of the Ca \textsc{ii}
8542 \AA\ line. Seeing induced cross-talk can appear given that all modulation states 
are observed at different times and reconstructed separately. The monochromatic images for 
all Stokes parameters from the NB camera are obtained after applying the calibrated 
modulation matrix for SST/CRISP. Our reconstructions and those
obtained with MOMFBD are very comparable. We find very similar signals in Stokes $V$ and Stokes $U$, 
although larger differences are found for Stokes $Q$. We also find some noise
in the umbra of the sunspot. We checked that this noise decreases as one increases $\lambda_\alpha$. 
Consequently, we associate the appearance of this noise due to the lack of contrast in the umbra, 
which make the estimation of the per-pixel wavefront coefficients in these regions uncertain. We
defer for the future the analysis of more elaborate regularization techniques to reduce these artifacts.
We expect that avoiding the separate reconstruction of all modulated images can help 
improve the polarimetric capabilities of multi-object multi-frame blind deconvolution.

\begin{figure*}
  \centering
  \includegraphics[width=0.9\textwidth]{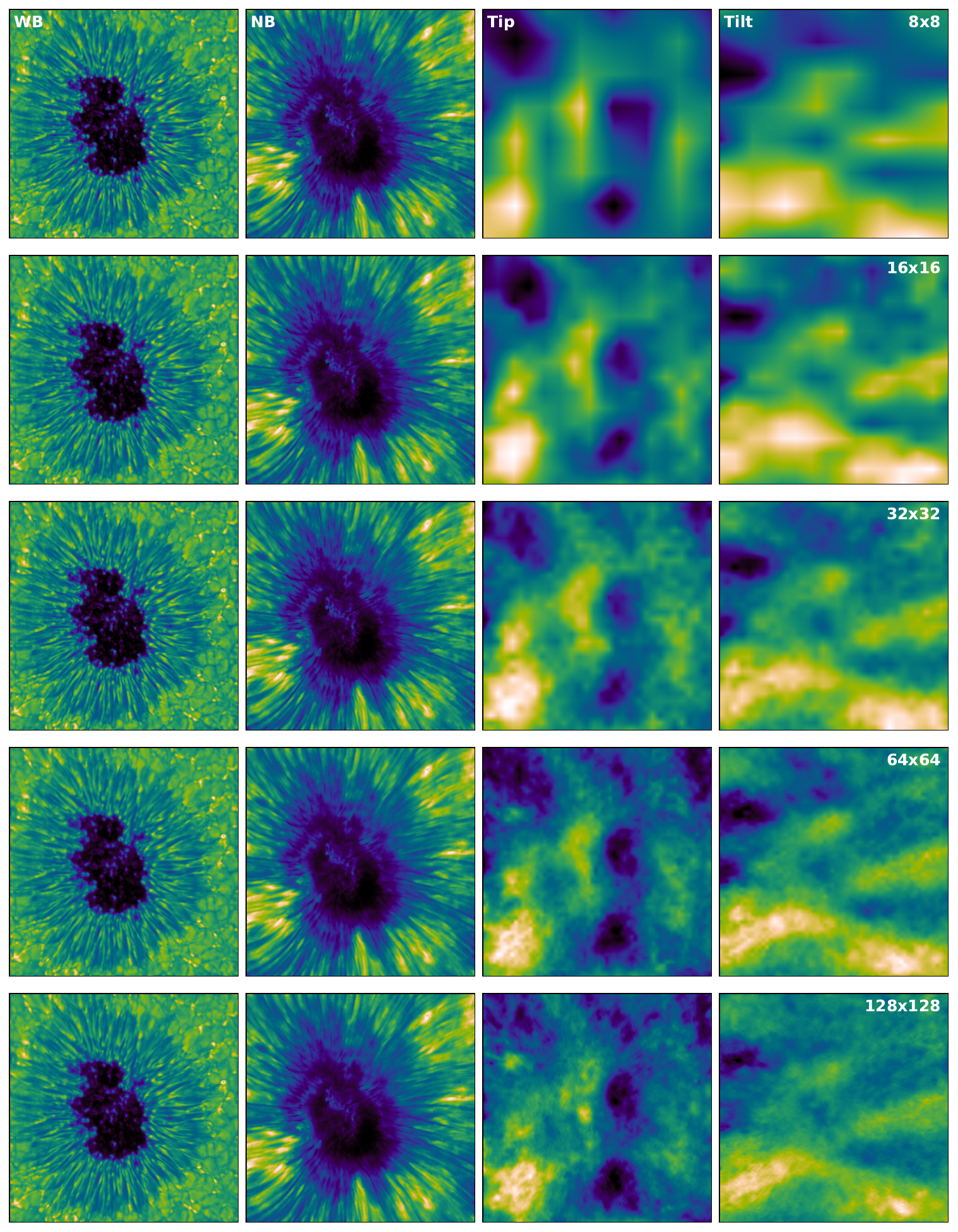}
  \caption{Reconstructions using grids of different number of pixels for the 
  KL modes. The first and second columns display the reconstructed WB and NB
  images, respectively. The third and fourth columns show the inferred tip and 
  tilts, respectively.}
  \label{fig:modes_npix}
\end{figure*}

As a final example, the SV-MOMFBD is used for reconstructing observations
from the HiFI instrument and the results are displayed in Fig. \ref{fig:hifi}.
The observations were acquired on 28 Nov 2022 at
position (226'', $-199$''), both in H$\alpha$ and the surrounding continuum. 
The left panel of Fig. \ref{fig:hifi} shows the first frame in the first column 
for the WB and NB channels. A burst of 50 frames 
taken at a cadence of 100 Hz and an exposure time of 9 ms with a pixel size of 0.050'' 
is used for the reconstruction. The high-altitude
seeing was not very good and quiet large spatially-variant tip-tilts are present
in the observed frames. For this reason, and given the limited tip-tilts that we 
used for training the SVCE, we decided to destretch the images similar to what
was done by \cite{2023SoPh..298...91A}. This can be solved in the future by using 
larger tip-tilts during the training of the SVCE. Our reconstructions (middle column) 
are compared with those obtained
with MOMFBD (right column) using 100 frames with 120 KL modes. Despite the comparison
is not direct, in our opinion, the results with 
120 KL modes seem to be a little over-reconstructed, with some artifacts visible in the WB images.
This specific reconstruction is obtained with $\lambda_o=0.005$ and $\lambda_\alpha=0.01$ and we used
a grid of $64 \times 64$ pixels for the modes. A Fourier filter with $w=1.85$ and $n=20$ is used.
The quality of the reconstruction depends on the specific hyperparameters used. 
Although the improvement with respect to individual frames is 
notable, the comparison is not as satisfactory as those found in \cite{2023SoPh..298...91A} 
and need to be improved in the future. Although the recovered WB contrast is similar to
that of MOMFBD, it is clear that we are missing some details, which are more conspicuous
in the NB channel. The right panel of Fig. \ref{fig:hifi} shows the azimuthally averaged
spatial power spectra for the WB channels. Both MOMFBD and SV-MOMFBD have a well-defined
cut close to the diffraction limit. However, MOMFBD recovers more power at large spatial
scales than SV-MOMFBD (between 0.04 px$^{-1}$ and 0.16 px$^{-1}$, rouhgly equivalent to the range between
6 and 23 px). We are actively working on improving the quality of the reconstructions for HiFI data.

\subsection{Modes}
Our SV-MOMFBD allows the user to select the grid where the 
modes are inferred. All previous experiments are carried out with a grid of 
64$\times$64. As already described, the per-pixel wavefront coefficients required
by the SV-MOMFBD are obtained by bilinear interpolation. It is interesting to 
analyze the impact of using a different grid for the modes. This is shown in 
Fig. \ref{fig:modes_npix}, where we show the spatial maps of $\alpha_1$ and
$\alpha_2$, the first two KL modes (tip and tilt),
from a grid of 8$\times$8 up to 128$\times$128. All calculations
are carried out with the same hyperparameter $\lambda_\alpha$. The first two columns 
show the reconstructed WB and NB images in each case, respectively. The
third and fourth columns show the tip and tilt KL modes for the first 
frame of the burst, respectively. We note that the inferred tip and tilt are 
very similar independently of the number of pixels in the grid. Obviously, those 
with a coarser grid are a coarse representation of those with a finer grid, where 
tinier details appear. Although the number of unknowns in a factor 256 larger in 
the finer grid, we do not find any hint of overfitting. Despite the large 
difference in the number of inferred modes, the reconstructions are all 
succesful. 

It is commonly assumed that it is better to reduce the number of unknowns when solving
an inverse problem. This would favor, in our case, using coarse grids. However, we instead
follow the philosophy of modeling everything (even if the number of parameters
increase much) so that we get close to the limit of inferring
almost per-pixel modes and use a much finer grid. With the appropriate 
regularization that we use (or any other regularization that we can 
propose in the future), we find no indication of overfitting. Additionally, a
significant amount of time is 
spent in the backpropagation through the bilinear interpolation operator when 
a coarse grid is used, so that finer grids give much faster reconstructions.

\section{Conclusions}
The post-facto correction of the perturbations introduced by the 
atmosphere when observing with large telescopes is a computationally 
challenging effort. The small anisoplanatic angles produce PSFs 
that vary strongly across the field of view. The most used code in the community,
MOMFBD, solves this problem by using the OLA method, where 
reconstructions are carried out in small patches where
the PSF is assumed to be constant (and the FFT can be used to 
carry out convolutions), which are finally stitched 
together. This approach, although successful, has several problems. The
first one is that it cannot be efficiently applied when anisoplanatic patches 
are very small (in very bad seeing conditions or very large telescopes).
The second one is that artifacts can appear when stitching together the 
patches. The MOMFBD code has dealt with this issue fantastically well and these 
artifacts are barely visible, except in a few exceptional cases. Third, 
carrying out convolutions using FFT in overlapping patches is not 
efficient.

In this paper we have shown that it is possible to train a deep
neural network to emulate convolutions with spatially variant PSFs.
The SVCE is able to produce sufficiently good convolutions so that it 
can be used as a forward model for image reconstruction. The SVCE is
fed with an image and a set of images representing the modes of the 
per-pixel wavefront represented in the KL basis. Although we train the model
using spatially invariant convolutions, we verify that the model generalizes
correctly to the spatially variant case. The model is very fast and 
fully convolutional, allowing it to be applied to images of arbitrary size
(although the input and output images are of the same size only when 
the input image has dimensions that are multiple of 2$^4$.) Apart
from the application shown in this paper, there are many other 
applications that can be explored in the future. One of them of
great relevance is the acceleration of simulations of multi-conjugate 
adaptive optics systems.

We have used the model for the solution of the multi-object, multi-frame 
blind deconvolution problem successfully. The ensuing code, SV-MOMFBD, is able 
to reconstruct large images at once, also inferring per-pixel PSF. The 
results displayed in this paper demonstrate that using an SVCE is 
an interesting avenue to accelerate image reconstruction for very large
telescopes. The reconstructions carried out in this paper have been carried 
out in standard desktop computers with off-the-shelf GPUs in a 
matter of tens of seconds for images as large as 512$\times$512. This is 
in opposition to the MOMFBD code, that requires the use of supercomputers.
We provide the code for training the SVCE, together with the trained
model used in this work, and the code to carry out the multi-object,
multi-frame blind deconvolution\footnote{\url{https://github.com/aasensio/spatially_variant}}.

A few caveats are in order. Although the SV-MOMFBD produces good 
reconstructions, they can be improved to remove artifacts. On one hand, the SVCE model needs to 
produce better convolutions. Larger training 
sets can be built, using more images similar to those that can be found
in the solar observations. A possibility to be explored is using 
synthetic observations obtained from radiation magneto-hydrodynamical
simulations, similar to the line followed by \cite{2019A&A...626A.102A}. 
Additionally, in order to produce 
more realistic convolutions, the U-Net model should be forced to produce better
results in the point-like objects. This can be reached by increasing the 
number of point-like objects in the training set or tweaking the U-Net 
architecture. We defer this study for the future.
On the other hand, the deconvolution implementation shown in this paper
lacks many of the technical improvements (phase diversity, pre-alignment of
images, removal of polarimetric cross-talk, \ldots) implemented in
the current version of the MOMFBD code \citep[see][for more details]{2021A&A...653A..68L}. Adding 
them will improve the quality of the reconstructions.

Another caveat is that we have only explored $\ell_2$ regularizations for
the smoothness of the reconstructed image and the modes. This has 
given us good results in many reconstructions, but not in all of
them. For instance, reconstructing off-limb structures can produce artifacts 
due to the lack of signal that are arguably better dealt with 
other type of regularizations. This needs to be explored.

A final caveat is that we have used the KL modes as representations
of the PSF. Given the presence of ambiguities between wavefronts and 
PSFs (the same PSF can be obtained with different wavefronts), this
representation might not be ideal. We want to explore  
factorizations extracted empirically from the PSF. An option is
to use principal components analysis to extract an orthogonal
basis for PSFs \citep[e.g.,][]{Lauer2002DeconvolutionWA}, or non-negative matrix factorization 
\citep{paatero1994positive,lee2001advances} to extract 
non-orthogonal basis.

The large GPU memory consumption during backpropagation turns out 
to be a problem for reconstructing very large 
images with our method. Although we have used some tricks such as working in 
half-precision and using gradient checkpointing, deconvolving 
very large bursts of many frames is not possible yet. Possible
solutions to this limitation are using multi-GPU architectures or
simply working with CPUs.

Finally, we have demonstrated that multi-object, multi-frame blind 
deconvolutions can be carried out with SV-MOMFBD using optimization. 
However, coupling our SVCE with architectures like those explored by \cite{AsensioRamos2020}
and \cite{2023SoPh..298...91A}
will probably produce new advances in speed. In such a case, we can build a larger model composed
of a neural network that infers the per-pixel modes 
from the burst of images and another
neural network that infers the reconstructed image. The output of both 
models can be fed into the SVCE and produce outputs that 
should be similar to the input burst. This large model can be 
trained unsupervisedly. This avenue is of great interest for the 
deconvolution of observations from solar telescopes of the 
4m class like the \emph{Daniel K. Inouye} Solar Telescope \citep[DKIST;][]{2020SoPh..295..172R} and the European Solar Telescope 
\citep[EST;][]{2022A&A...666A..21Q}.

\begin{acknowledgements}
We thank Mats L\"ofdahl for useful discussions.
We acknowledge support from the Agencia Estatal de Investigaci\'on del 
Ministerio de Ciencia, Innovaci\'on y Universidades (MCIU/AEI) 
and the European Regional Development Fund (ERDF) through project PID2022-136563NB-I0.
This research has made use of NASA's Astrophysics Data System Bibliographic Services.
We acknowledge the community effort devoted to the development of the following 
open-source packages that were
used in this work: \texttt{numpy} \citep[\texttt{numpy.org},][]{numpy20}, 
\texttt{matplotlib} \citep[\texttt{matplotlib.org},][]{matplotlib}, and \texttt{PyTorch} 
\citep[\texttt{pytorch.org},][]{pytorch19}.
\end{acknowledgements}

%
\bibliographystyle{aa} 

%

\end{document}